\documentclass[a4paper,11pt]{article}
\pdfoutput=1

\usepackage{jheppub}
\usepackage[T1]{fontenc}
\usepackage{amssymb}
\usepackage{graphicx}
\usepackage{amsmath}
\usepackage{tensor}
\usepackage{hyperref}
\usepackage{epstopdf}
\usepackage{extarrows}
\usepackage{xcolor}
\usepackage{wrapfig,boxedminipage,setspace,subfigure,epsfig}
\usepackage{tikz}
\usetikzlibrary{decorations.pathmorphing,calc}

\newcommand{\td}{\text{d}}

\newcommand{\Tr}{\text{Tr}}

\begin{document}

\title{Holographic entanglement of purification for thermofield double states and thermal quench}

\author[a]{Run-Qiu Yang}
\author[b]{Cheng-Yong Zhang,}
\author[c]{Wen-Ming Li}

\emailAdd{aqiu@kias.re.kr}
\emailAdd{zhangchengyong@fudan.edu.cn}
\emailAdd{liwmpku@pku.edu.cn}

\affiliation[a]{Quantum Universe Center, Korea Institute for Advanced Study, Seoul 130-722, Korea}
\affiliation[b]{Department of Physics and Center for Field Theory and Particle Physics, Fudan University, Shanghai 200433, China}
\affiliation[c]{Department of Physics and State Key Laboratory of Nuclear Physics and Technology, Peking
University, No.5 Yiheyuan Rd, Beijing 100871, P.R. China}

\abstract{
We explore the properties of holographic entanglement of purification (EoP) for two disjoint strips in the Schwarzschild-AdS  black brane and the Vaidya-AdS black brane spacetimes. For two given  strips  
on the same boundary of Schwarzschild-AdS  spacetime,
there is an upper bound of the separation beyond which the holographic EoP will always vanish no matter how wide the strips are. In the case that two strips are in the two boundaries of the spacetime respectively, we find that the holographic EoP exists only when the strips are wide enough. If the width is finite, the EoP can be nonzero in a finite time region. For thermal quench case, we find that the equilibrium time of holographic EoP is only sensitive to the width of strips, while that of the holographic mutual information is sensitive not only to the width of strips but also to their separation.
}

\maketitle
\flushbottom

\noindent

\section{Introduction}
Entanglement is one of the most significant features of quantum physics,
and plays an important role in understanding quantum many-body physics,
quantum field theory, quantum information as well as quantum gravity.
In quantum field theory, the entanglement entropy (EE) measures the
entanglement between a subregion $A$ of Hilbert space and its complement
$\bar{A}$.  It is defined as the von Neumann entropy of the reduced
density matrix,
\begin{equation}
S_{A}:=-\text{tr}\rho_{A}\log\rho_{A}\label{eq:EE}
\end{equation}
where $\rho_{A}:=\text{tr}_{\bar{A}}\rho$ is the reduced density matrix of
$A$ with respect to the density matrix of the whole system. In the
AdS/CFT correspondence\cite{Maldacena:1997re}, there is a simple holographic
counterpart given by the Hubeny-Rangamani-Takayanagi (HRT) surface \cite{PhysRevLett.96.181602,Ryu:2006ef},

\begin{equation}
S_{A}=\frac{\text{Area}(\gamma_{A})}{4G_{N}}.\label{eq:RT formula}
\end{equation}
where $G_{N}$ is the Newton constant, of which the relation with
the central charge of CFT is $c=\frac{3}{2G_{N}}$, and $\gamma_{A}$ is
the extremal surface sharing the common boundary with $A$ and is homologous
to $A$. In this paper, we will set $G_N=1$.

For pure state, the EE computed by Eq.~\eqref{eq:EE} is the only way to characterize the quantum entanglement of a given bipartite system.  However, when the system is in a mixed state it is not. There are several different quantities to describe the quantum or classical correlations between two subsystems $A$ and $B$. For example, one of the well-studied quantity both in quantum information theory and its holographic duality is the mutual information (MI) $I(A:B)$~\cite{PhysRevE.69.066138,Fischler:2012uv,Morrison:2012iz}, which is defined as
\begin{equation}\label{defmutualI}
  I(A:B)=S(\rho_A)+S(\rho_B)-S(\rho_{AB}),
\end{equation}
where $AB=A\cup B$. But this quantity is only the linear combination of EE, so it is not a new quantity both in the view point of holographic duality or quantum information theory.

Recently, a new quantity describing the entanglement between mix states, the entanglement
of purification, was studied in holographic duality \cite{Takayanagi:2017knl}. Entanglement
of purification (EoP) \cite{doi:10.1063/1.1498001} is defined by minimum
EE for all possible purification of the mixed state, which is defined
as
\begin{equation}
E_{P}(A:B)=\min_{\rho_{AB}=Tr_{A'B'}|\Psi\rangle\langle\Psi|}S(\rho_{AA'}).\label{eq:EoP}
\end{equation}
Here $|\Psi\rangle$ is a pure state on the enlarged Hilbert space
$\mathcal{H}_{A}\otimes\mathcal{H}_{B}\otimes\mathcal{H}_{A'}\otimes\mathcal{H}_{B'}$,
where $\mathcal{H}_{A}\otimes\mathcal{H}_{B}$ is the initial Hilbert
space in which the mixed state $\rho_{AB}$ lives, and $\mathcal{H}_{A'}$
(or $\mathcal{H}_{B'}$) is arbitrary that is needed in order to purify
the mixed state. The EoP can be viewed as a generalization of EE, as evidently
for the pure state it equals  EE. The EoP of bipartite system is zero only when $\rho_{AB}=\rho_A\otimes\rho_B$.

The EoP has strong relationship to MI. In fact, we have~\cite{PhysRevA.91.042323}
\begin{equation}\label{ineqEoPS}
  \frac12I(A:B)\leq E_P(A:B)\leq\min\{S(\rho_A),S(\rho_B)]\,.
\end{equation}
This inequality is saturated in both sides if $AB$ is a pure state. When the MI vanishes\footnote{The MI of two disjoint regions is usually nonvanishing due to the quantum correlations between them, see \cite{Wolf:2007tdq,Chen:2017hbk,Chen:2017yns} for example. However, we work in the classical gravity limit in this paper.},
 the bipartite system $AB$ is separable and so  we have $E_P(A:B)=0$. Thus, the nonzero EoP can appear only when MI is nonzero.

Since there are infinite ways
of purification, it is hard to work out the EoP in the CFT side \cite{Bhattacharyya:2018sbw,Hirai:2018jwy}
(early works focused on spin systems in numerical such as \cite{2017arXiv171101288H,2018arXiv180100142C}).
In  recent works, inspired by the RT formula, a holographic formula
for the EoP was proposed in \cite{Takayanagi:2017knl,Nguyen:2017yqw}
and generalized to multipartite and other situations in \cite{Bao:2017nhh,Bao:2018gck,Espindola:2018ozt,Umemoto:2018jpc,Tamaoka:2018ned}.
In this holographic conjecture, the EoP is dual to the entanglement
wedge \cite{Headrick:2014cta} cross section $E_{W}$, which reads
\begin{equation}
E_{P}=E_{W}.\label{eq:EpEw}
\end{equation}
This conjecture is powerful since it implies that the holographic
state dual to the surface of entanglement wedge is an optimal purification
of the density matrix of any geometric subregion of the boundary theory.
Evidently, when the state is a pure one, $E_{W}$ is equal to the
EE, which is the same as that in the CFT side.

Now for a bipartite mixed state, we have three different quantities with their holographic descriptions at hand, i.e., the EEs of $A$ and $B$, the mutual information  $I(A:B)$ and the entanglement of purification $E_P(A:B)$. The former two have been studied deeply both from information theories and holographic duality. However, the behavior of EoP is not known well at current. Though a few of works have been done to understand EoP and its holographic duality from conformal field theory~\cite{Hirai:2018jwy,Bao:2017nhh,Espindola:2018ozt,Umemoto:2018jpc}, it is not known well how different the EoP will be if when we compare it with EE and MI in some concentrate systems. It is also important to find what new properties can be carried by EoP when we study the entanglement between two subregions in a mixed state. The main aim of this paper is to make some preliminary explorations on these aspects.

In this work, we will explore the properties of holographic EoP in the Schwarzschild-AdS  black brane and the Vaidya-AdS black brane, which
are dual to the thermofield double state and thermal quench respectively. We first consider two disjoint strips with the same width on the same boundary of Schwarzschild-AdS  black brane, which is dual to the two disjoint subregions of a thermal state.
 In the case that two strips are in two boundaries of the extended black brane respectively, we consider how the EoP evolves according to the boundary time.
 Finally we also consider the EoP in the quench case.
 Since the holographic EoP exists only when the holographic MI is positive, we will further compare the evolution behaviors of holographic MI and EoP.

The organization of this is as follows. In Sec.~\ref{EoPHolo1}, we first consider the two disconnected regions in the same side and try to discover the relation between EoP and the size of the subregion, the separation of the subregion. In Sec.~\ref{EoPHolo2}, we consider the time evolution of EoP when the two subregions in different sides. We will consider how the widths of two regions effect time-evolutional behaviour of EoP. In Sec.~\ref{quenchEoP}, we will study the effects of thermal quench on EoP of two regions by Vaidya-AdS black brane. A short summary will be found in Sec.~\ref{sum}.

\section{Entanglement of purification for two strips on the same side}\label{EoPHolo1}
We consider the Schwarzschild AdS black brane in $(d+1)$-dimensional case. The metric reads,
\begin{equation}\label{metricAdS}
  \td s^2=\frac1{z^2}\left[-f(z)\td t^2+\frac{\td z^2}{f(z)}+\td\vec{x}^2_{d-1}\right],~~~f(x):=1-z^d/z_h^d\,.
\end{equation}
Here $\td\vec{x}^2_{d-1}$ is induced line elements at the spatial $(d-1)$-dimensional subspace with $z=0$ and constant $t$. The spatial coordinates are $\{x_1,x_2,\cdots,x_{d-1}\}$.   $z_h$ is the inverse horizon radius and the inverse temperature of dual boundary theory is $\beta=4\pi z_h/d$. In this paper, we  set $z_h=1$ so the inverse temperature $\beta=4\pi/d$.

\begin{figure}
\begin{centering}
\includegraphics[scale=0.5]{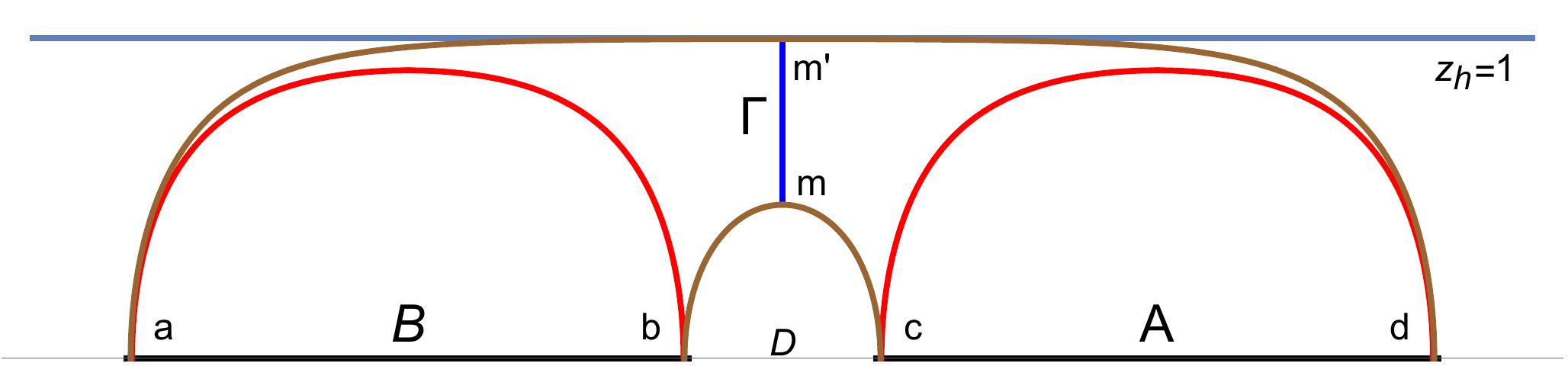}
\par\end{centering}
\caption{\label{fig:The-finite-strip}The finite strips on the same boundary of a time slice of the Schwarzschild-AdS black brane spacetime. $m$ and $m'$ are two turning points of minimal surface connecting $ad$ and $bc$. $\Gamma$ is the
cross section of entanglement wedge when the entanglement wedge is connected.}
\end{figure}

Let us first consider the case that the subregions $A$ and $B$ are both infinite strips separated by distance $D$ on the same boundary of the spacetime at fixed time $t=0$ (see Fig. \ref{fig:The-finite-strip}).
The subregions are
$$A:=\{l+D/2>x_1>D/2, -\infty<x_i<\infty,i=2,3,\cdots,d-1\}$$
and
$$B:=\{-l-D/2<x_1<-D/2, -\infty<x_i<\infty,i=2,3,\cdots,d-1\}\,,$$
%

As discussed in  ~\cite{Takayanagi:2017knl}, when the two subsystems are separated from each other far away enough, the system is the product state of $A$ and $B$, the entanglement wedges are disconnected and so there is no holographic EoP. The transition point of nonzero EoP can be find by the inequality  ~\eqref{ineqEoPS}.
For a strip with width $w$, the holographic entanglement entropy is
\begin{equation}
S(w)=\frac{2V_{d-2}}{4 }\int_{\delta}^{z_{m}}\frac{dz}{z^{d-1}}\frac{1}{\sqrt{\left(1-z^{d}\right)\left(1-\frac{z^{2d-2}}{z_{m}^{2d-2}}\right)}}\,,\label{eq:OneSideSz}
\end{equation}
where $V_{d-2}:=\int\td x_2\cdots\td x_{d-1}$ and $z_{m}$ is the  turning point
of the minimal surface corresponding to the strip with width $w$,
of which their relation is given by
\begin{equation}
w=2\int_{\delta}^{z_{m}}dz\frac{1}{\sqrt{\left(1-z^{d}\right)\left(\frac{z_{m}^{2d-2}}{z^{2d-2}}-1\right)}}.\label{eq:OneSideLz}
\end{equation}
From Fig.~\ref{fig:The-finite-strip}, we can see that $S_A=S_B=S(l)$ and $S_{AB}=S(2l+D)+S(D)$. Thus the holopgraphic MI of $AB$, which is the function of $D$ and $l$, can be expressed as
\begin{equation}\label{mutinfAB1}
I(D,l)=S_A+S_B-S_{AB}=2S(l)-S(D)-S(2l+D).
\end{equation}
The holographic EoP is nonzero only if $I(D,l)>0$.
%
\begin{figure}
\begin{centering}
\begin{tabular}{cc}
\includegraphics[scale=0.44]{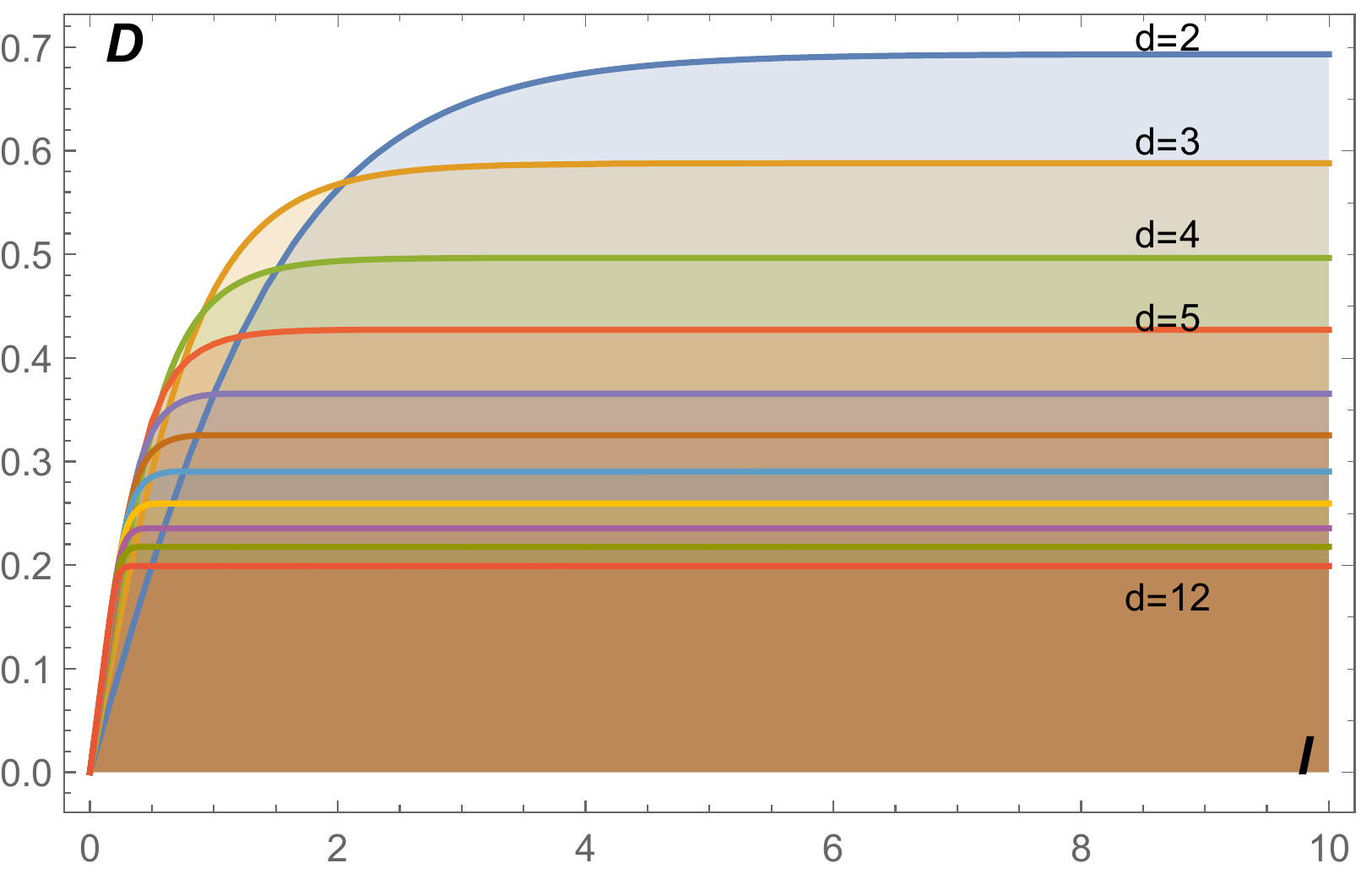} & \includegraphics[scale=0.44]{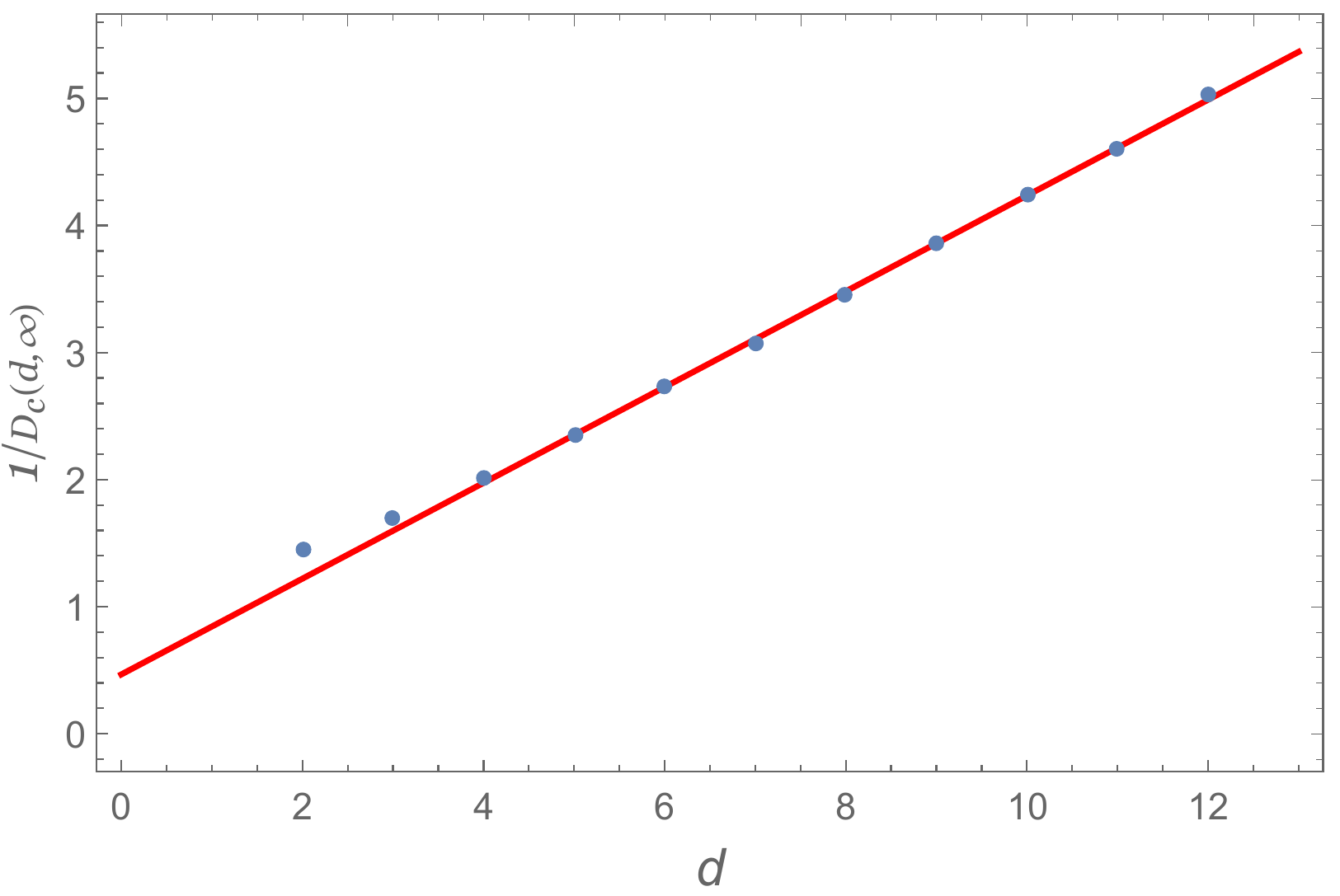}\tabularnewline
\end{tabular}
\par\end{centering}
\caption{\label{fig:FiniteCriticalLentgth}Left panel: The regions below the
lines are allowed to have non-vanishing holographic EoP in different dimensional spacetimes.
Right panel: The critical length $D_{c}$ of separation when
$l\to\infty$ in different dimensions.}
\end{figure}
The regions having EoP for different dimensional spacetimes are shown
in the left panel of Fig.~\ref{fig:FiniteCriticalLentgth}. For given dimension $d$ and strip width $l$, there is a critical separation $D_c(d,l)$ and the holographic EoP is nonzero only if $D<D_c(d,l)$.
When $d=2$, we can work out the critical separation
$D_{c}(2,l)$ analytically,
\begin{equation}\label{DLBTZ1}
\cosh\frac{D_c(2,l)}{2}=\sqrt{1+2\sqrt{2\cosh l}\cosh\frac{l}{2}+2\cosh l}\left[\cosh\frac{3l}{2}-\sqrt{2}(\cosh l)^{3/2}\right].
\end{equation}
 The critical separation when $l\to\infty$ is $D_{c}(2,\infty)=\ln2=\frac{\beta}{2\pi}\ln2$.
This means that when $D>\ln2$, there is no holographic EoP no matter how large $l$ is.
When $d>2$, one can work out the critical separation numerically. The results are shown in Fig. \ref{fig:FiniteCriticalLentgth}. From the left panel, we see that for small given $l$, $D_c(d,l)$ grows with $d$. For larger given $l$, $D_c(d,l)$ decreases with $d$.
%
For given $d$, when $l$ is small, the critical separation
grows linearly with $l$:
\begin{equation}\label{smalllDl1}
  D_c(d,l)\simeq c_{0}(d)l\,.
\end{equation}
When $d=2$, the coefficient $c_{0}(2)=\sqrt{2}-1.$ For larger $d$, $c_{0}(d)\simeq\frac{1}{1.8-1.4d}+1$  approximately.
The critical separation
$D_{c}(d,l)$ when $l\to\infty$ is related approximately to the spacetime dimension $d$  by $
D_{c}(d,\infty)^{-1}\simeq0.5+0.4d.$
For large enough $l$, the critical separation tends to $D_{c}(d,\infty)$ asymptotically and there are $D_{c}(d,l)\simeq D_{c}(d,\infty)-c_{1}(d) e^{-c_{2}(d)l}$ where $c_{1}(d),c_{2}(d)$ are positive constants depending on $d$.

\begin{figure}
\begin{centering}
\begin{tabular}{ccc}
\includegraphics[scale=0.28]{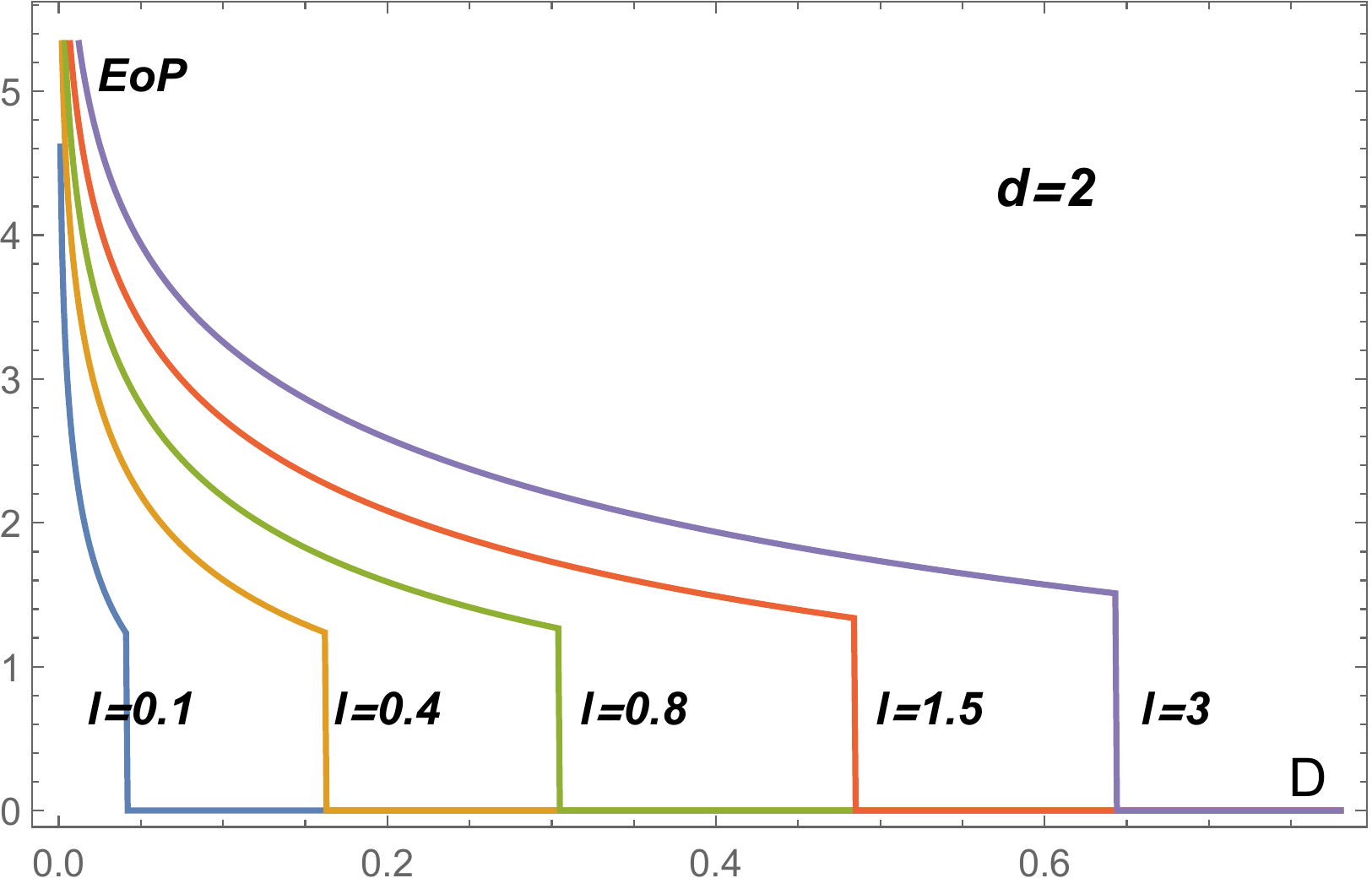} & \includegraphics[scale=0.28]{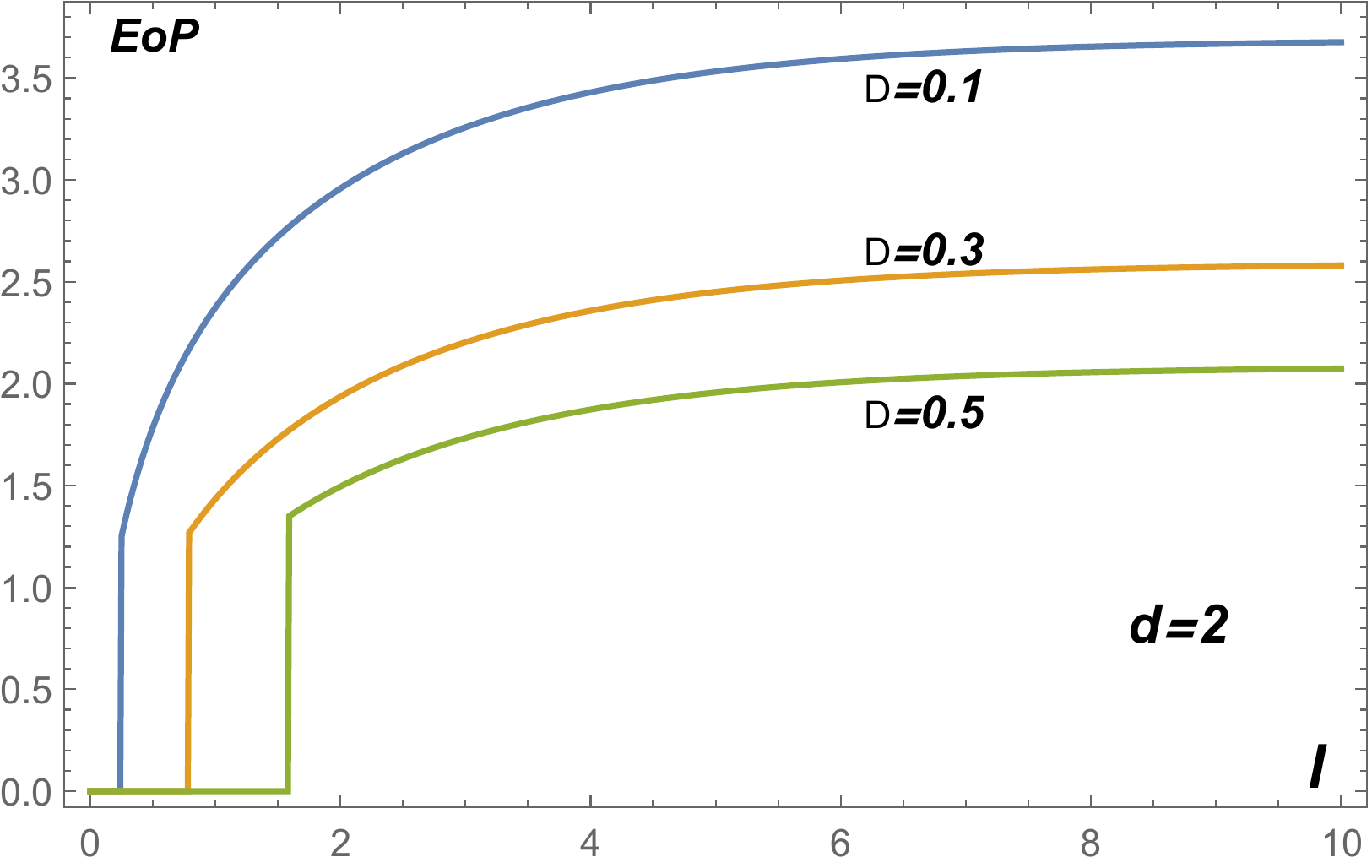} & \includegraphics[scale=0.28]{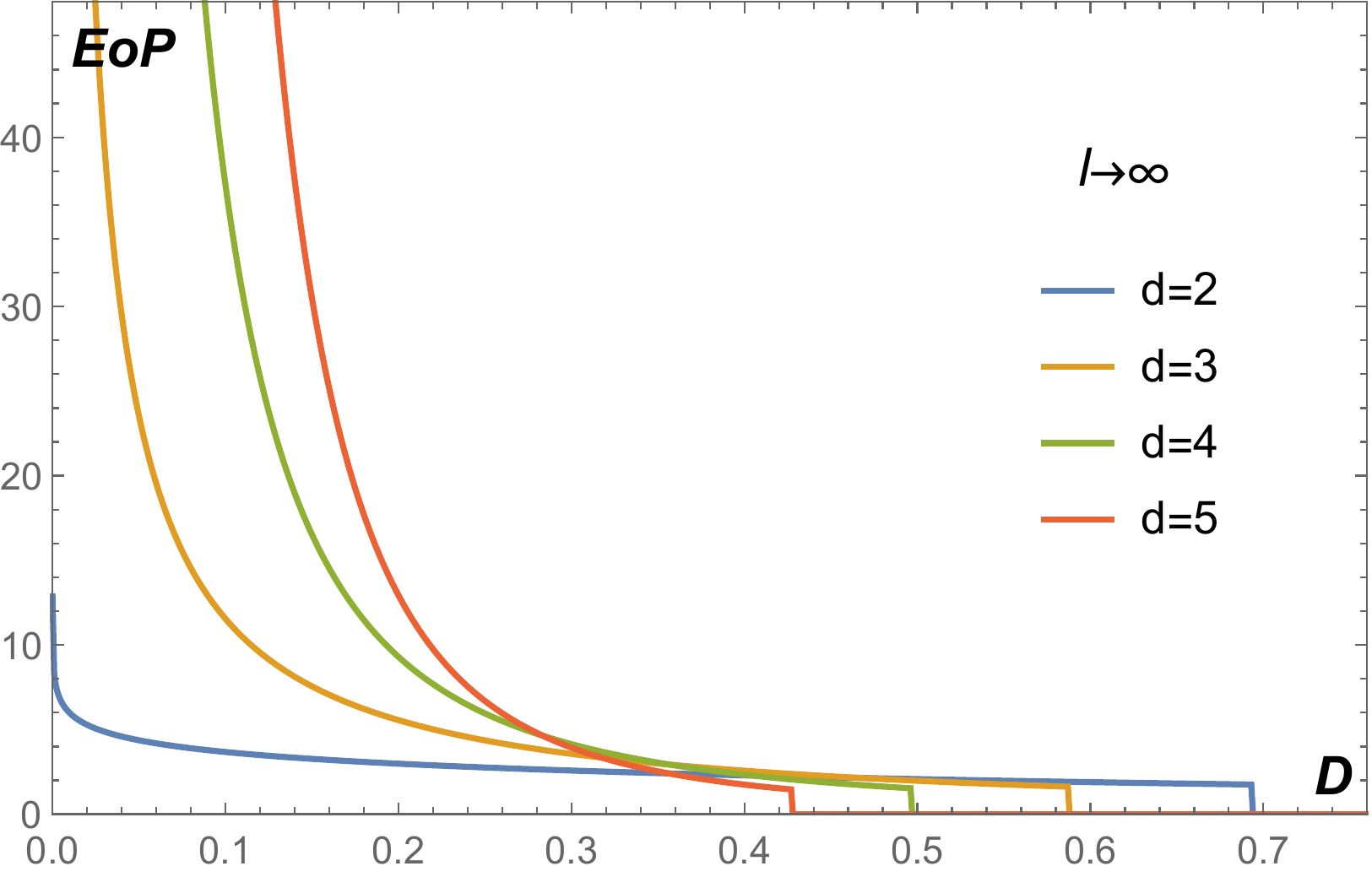}\tabularnewline
\end{tabular}
\par\end{centering}
\caption{\label{fig:LFiniteEoP}Left and middle panels: The EoP (in unit of $4/V_{d-2}$) for different
$l$ and $D$ when $d=2$. Right panel: EoP for strip with very large $l$ in different dimensional spacetimes. }
\end{figure}

When $I(D,l)>0$, the holographic EoP is   given by
\begin{equation}\label{Eforoneside}
\frac{4 }{V_{d-2}}E(l,D)=\begin{cases}
\ln\frac{\tanh(\frac{D+2l}{4})}{\tanh(\frac{D}{4})}, & d=2,\\
\left.\frac{-4z^{2-d}\sqrt{1-z^{d}}+(d-4)z^{2}F\left(\frac{1}{2},\frac{2}{d},\frac{2+d}{d},z^{d}\right)}{4(d-2)}\right|_{z_{D}}^{z_{2l+D}}, & d>2.
\end{cases}
\end{equation}
 We plot the holographic EoP for different $l$ and $D$ when $d=2$ in the
left and middle panels of Fig. $\ref{fig:LFiniteEoP}$. In the left panel, we see that when the separation $D$
goes to zero,  the holographic  EoP goes to infinity. This is due to the UV divergence
near the spacetime boundary. As $D$ grows, the holographic  EoP takes a nosedive.
The change becomes slowly as $D$ grows further. However, when
$D$ goes beyond the upper bound, i.e., $D>D_c(d,l)$, the holographic EoP drops suddenly to zero.
These are two phases corresponding to the connected entanglement wedge and
the disconnected one, respectively. Moreover, the smaller the strip width $l$ is, the shorter $D$ having holographic EoP is and the sharper the nosedive is.
For fixed separation, the holographic EoP vanishes when the strip width is small, as shown in the middle panel. It becomes positive discontinuously when the strips are wide enough. When the strip width is very large, the holographic EoP tends to a saturation value. The larger the separation is, the smaller saturation of holographic EoP.
In the right panel of Fig. $\ref{fig:LFiniteEoP}$, we show the holographic EoP in different dimensional spacetimes  for strips with $l\to\infty$. The holographic EoP decays slower with separation  in higher dimension. Beyond the critical separations, the holographic EoP drops discontinuously to zero.

\section{Entanglement of purification for two-side subregions}\label{EoPHolo2}
The maximally extended Penrose diagram of static black brane contains two-copy of the boundaries, which corresponds to two copies of the field theory. The full spacetime is conjectured dual to the thermofield double states~\cite{Maldacena:2001kr}. In general, these two copies are in an entangled state in the form
\begin{equation}\label{defTFD}
  |\Phi\rangle:=\frac1Z\sum_{n}e^{-\beta E_n/2}|E_n\rangle_L|E_n\rangle_R\,.
\end{equation}
The states $|E_\alpha\rangle_L$ and $|E_\alpha\rangle_R$ are eigenstates in the two copy field theories, $Z$ is the normalized factor and $\beta$ is the temperature of these two field theories.

In the last section, we only studied the subregions in one boundary state of the black branes.  In this section, we will consider the case that subregions $A$ and $B$ locate at the two boundaries, respectively. The union of two time slices $t_L=t_R=0$ at two boundaries is dual to a TFD state $|\Phi\rangle$ in ~\eqref{defTFD}. With the Hamiltonians $H_L$ and $H_R$ at the left and right dual CFTs, respectively, the time evolution of a TFD state then reads,\footnote{Here we choose the total Hamilton to be $H_L+H_R$. Alternatively, we can define the total Hamilton is $H_L-H_R$, by which the TFD state will not evolve with respective to boundary $t_B$. }
\begin{equation}\label{timesate1}
  |\Phi(t)\rangle:=e^{-it(H_L+H_R)}|\Phi\rangle\,
\end{equation}
This time-dependent TFD state can be characterized by the codimension-2 surfaces of $t_B=t_L=t_R$ at the two boundaries of the AdS black brane.

\subsection{Infinite size case}
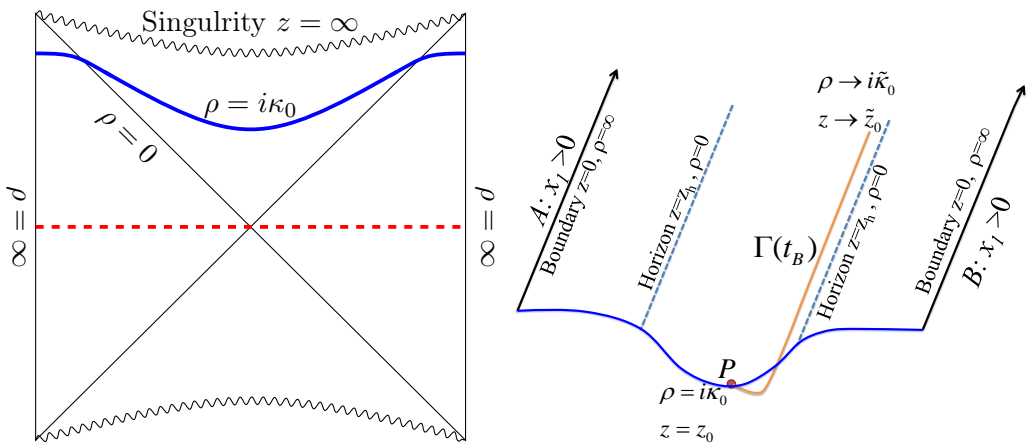
\begin{figure}	
	\begin{center}	
		\begin{tikzpicture}
		
		\node (I)    at (0,0)   {};
		
		\path 
		(I) +(135:4)  coordinate  (Iltop)
		+(-135:4) coordinate                     (Ilbot)
		+(0:0)   coordinate                  (Ilmid)
		+(45:4) coordinate        (Irtop)
		+(-45:4)   coordinate                  (Irbot)
		+(0:0)   coordinate                  (Irmid)
		;
		
		\draw  (Iltop) --
		node[pos=0.5, below, sloped]    {$\rho=\infty$}
		(Ilbot) --
		(Ilmid) --
		(Irtop) --
		node[pos=0.5, above, sloped]    {$\rho=\infty$}
		(Irbot) --
		node[midway, below, sloped]    {}
		(Irmid) --
		node[midway, below, sloped]    {$\rho=0$}
		(Iltop) -- cycle;
		
		\draw[decorate,decoration={snake,segment length=0.15cm,amplitude=0.05cm}] (Iltop)
		to[out=-15,in=+195,looseness=1.2]
				node[midway, above, inner sep=2mm] {Singulrity $z=\infty$}
		(Irtop);
		
		\draw[decorate,decoration={snake,segment length=0.15cm,amplitude=0.05cm}] (Ilbot)
		to[out=15,in=-195,looseness=1.2]
		(Irbot);
		
		\draw [blue,line width=.5mm]
		(-2.828,2.3)
		to[out=0,in=150,looseness=1.3]
		(-1.828,2);
		\draw [blue,line width=.5mm]
		(2.828,2.3)
		to[out=180,in=30,looseness=1.3]
		(1.828,2);
		\draw [blue,line width=.5mm]
		(-1.928,2.06) coordinate
		to[out=-30,in=210,looseness=1.36]
        node[black,midway, above]    {$\rho=i\kappa_0$}
		(1.928,2.06)  coordinate;



		\draw [dashed,red,line width=.5mm]
		(-2.828,0) -- (2.828,0);

		\end{tikzpicture}
\includegraphics[width=.45\textwidth]{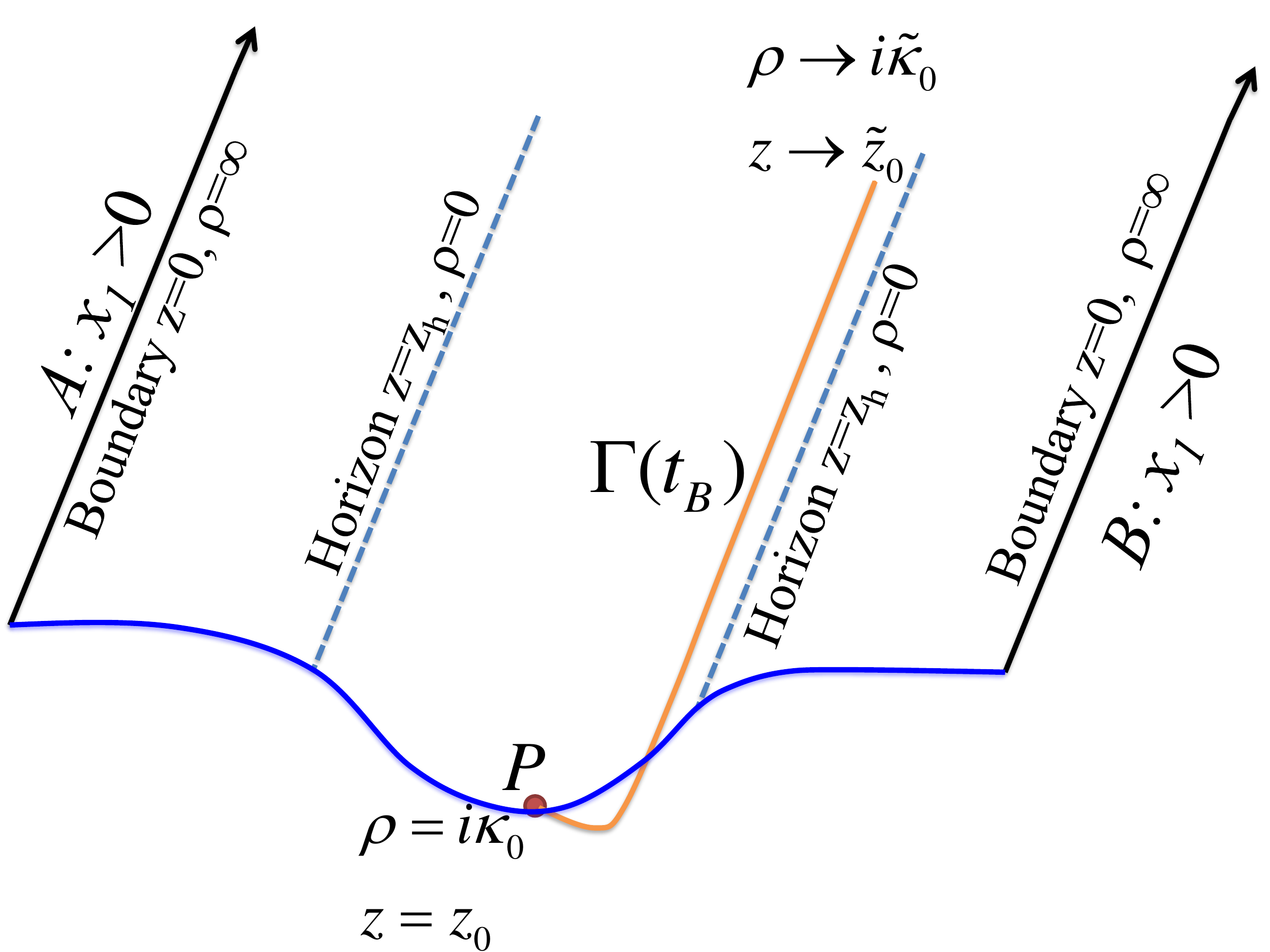}
		\caption{Extremal surfaces in the AdS black brane.  The two time slices at left and right boundary are given by $t_L=t_R=t$ and $\rho=\infty$. The half infinite subregions $A$ and $B$ locate at left and right boundaries, respectively.  The entanglement wedge cross section denoted by $\Gamma(t_B)$ hides in the inner region of the black brane. }\label{FigAdS}
	\end{center}
\end{figure}

Let us first consider the two infinitely wide strips appearing in the two boundaries symmetrically. The subregions are (see the right panel of Fig.~\ref{FigAdS})
$$A:=\{t=t_B, x_1>0, -\infty<x_i<\infty,i=2,3,\cdots,d-1\}$$
and
$$B:=\{t=t_B, x_1>0, -\infty<x_i<\infty,i=2,3,\cdots,d-1\}\,.$$
The induced density matrix of $A\cup B$ is also time-dependent,
\begin{equation}\label{inducedAB1}
  \rho_{AB}=\Tr_{L,x_1>0}\Tr_{R,x_1>0}(|\Phi(t_B)\rangle\langle\Phi(t_B)|).
\end{equation}
Thus, the EoP between $A$ and $B$ is also time-dependent. The union of $A\cup B$ and $\overline{A\cup B}$ gives the whole boundaries, so we can also study the entanglement entropy between  $A\cup B$ and $\overline{A\cup B}$, which is given by
\begin{equation}\label{defSABTFD}
  S_{AB}:=-\Tr(\rho_{AB}\ln \rho_{AB})\,.
\end{equation}
Now let us study $E(t_B):=E_P(A:B)$ and $S_{AB}$ by holographic duality.


%
%

In order to compute area $\Gamma$ in this case, we need first to find the boundary of entanglement wedge, i.e., the extremal surface connect $\partial A$ and $\partial B$ (see blue line in Fig.~\ref{FigAdS}). We rewrite the metric~\eqref{metricAdS} into following form
\begin{equation}
ds^{2}=-g^{2}(\rho)\td t^{2}+\td\rho^{2}+h^{2}(\rho)\td\vec{x}_{d-1}^{2}\label{eq:SAdSplanarBB},
\end{equation}
where
\begin{equation}
h(\rho)=\frac{2}{d}\left(\cosh\frac{d\rho}{2}\right)^{2/d},\ \ g(\rho)=h(\rho)\tanh\frac{d\rho}{2}.
\end{equation}
This metric is obtained after a coordinate transformation
$\td\rho=\frac{\td z}{z\sqrt{1-z^{d}}}$ and
we have set $z_h=1$ and so $\beta=4\pi/d$.
The Penrose diagram and the entanglement wedge cross section are shown in the Fig.~\ref{FigAdS}.

We can continue (\ref{eq:SAdSplanarBB}) into the interior region of Fig.~\ref{FigAdS} by setting $\rho=i\kappa$ and the replacement $t\rightarrow t+i\pi/2$. For the case $t_B\equiv t_R=t_L$, the maximal volume surface is given by the blue line in Fig.~\ref{FigAdS}. The red dotted line is for $t_B=0$. The corresponding codimension-two surface is obtained by extremalizing following integration
\begin{align}\label{eq:vol}
	\int h(\rho)^{d-2}\sqrt{-g^{2}(\rho)+(\partial\rho/\partial t)^{2}}\td t.
\end{align}
In principle, we should solve the Euler-Lagrangian equation of \eqref{eq:vol} to find
$\rho(t)$
. However, because the ``Lagrangian'' in ~\eqref{eq:vol} does not contain ``time'' explicitly, there is a ``conserved charge'' by which we can simplify the process to find the extremal surface.
Following \cite{Hartman:2013qma,MIyaji:2015mia,Sinamuli:2016rms} we may find the first integral of the equation of motion of \eqref{eq:vol},
which yields
\begin{align}\label{eq:energy}
\frac{g^2 h^{d-2}}{\sqrt{-g^2+(\partial\rho/\partial t)^{2}}}= & i g_0 h_0^{d-2}\,,
\end{align}
where $h_0 := h(i\kappa_0)$ and $g_0:= g(i\kappa_0)$
with $\kappa_{0}$ ($0<\kappa_{0}<\frac{\pi}{2d}$) satisfying $\frac{\partial\kappa}{\partial \tilde{t}}| _{\kappa=\kappa_0}=0$.
From (\ref{eq:energy}), we can write the time $t_B$ in terms of $\kappa_0$:
\begin{equation}\label{timekappa}
	\begin{split}	t_B=&\int_{\delta}^{\kappa_{0}}\frac{\td\kappa}{\left(\cos\frac{d\kappa}{2}\right)^{\frac{2}{d}}\tan\frac{d\kappa}{2}\sqrt{1-\frac{\cos^{\frac4d}(d\kappa_0/2)}{\cos^{\frac4d}(d\kappa/2)}\frac{\sin^{2}d\kappa}{\sin^{2}d\kappa_{0}}}}\\ &-\int_{\delta}^{\infty}\frac{\td\rho}{\left(\cosh\frac{d\rho}{2}\right)^{\frac{2}{d}}\tanh\frac{d\rho}{2}\sqrt{1+\frac{\cos^{\frac4d}(d\kappa_0/2)}{\cosh^{\frac4d}(d\kappa/2)}\frac{\sinh^{2}d\rho}{\sin^{2}d\kappa_{0}}}}\,.
	\end{split}
\end{equation}
Here we have introduced the IR cut off $\delta\rightarrow0$.
Substituting (\ref{eq:energy}) into (\ref{eq:vol}), the extremal volume can be expressed in terms of the parameter $\kappa_{0}$,
\begin{equation}\label{ETFDs1}
\begin{split}
	S_{AB}(t_B)=&2S_{\text{div}}+2V_{d-2} \left[\int_{0}^{\kappa_{0}}\frac{\left(\cos\frac{d\kappa}{2}\right)^{\frac{2(d-2)}{d}}}{\sqrt{\frac{\cos^{\frac4d}(d\kappa/2)}{\cos^{\frac4d}(d\kappa_0/2)}\frac{\sin^{2}d\kappa_{0}}{\sin^{2}d\kappa}-1}}\td\kappa\right.\\
&
	+\left.\int_{0}^{\infty}\left(\frac{\left(\cosh\frac{d\rho}{2}\right)^{\frac{2(d-2)}{d}}}{\sqrt{1+\frac{\cosh^{\frac4d}(d\kappa/2)}{\cos^{\frac4d}(d\kappa_0/2)}\frac{\sin^{2}d\kappa_{0}}{\sinh^{2}d\rho}}} -\frac{\cosh\frac{d\rho}{2}}{\left(\sinh\frac{d\rho}{2}\right)^{\frac{4-d}{d}}}\right)\td\rho
		\right].
\end{split}
\end{equation}
Here $S_{\text{div}}$ is the universal UV divergent term for extremal surface, which reads
\begin{equation}\label{defUVS}
  S_{\text{div}}:=\left\{
  \begin{split}
  &\frac14\ln(\beta/\pi\epsilon),~~~d=2\\
  &\frac{V_{d-2}}{4(d-2)}\left(\frac{\beta d}{4\pi\epsilon}\right)^{d-2},~~d>2\,.
  \end{split}
  \right.
\end{equation}
%
In the case of $\kappa_0\rightarrow\frac2d\arcsin\sqrt{\frac{d}{2d-2}}$,
the boundary time $t_B\rightarrow\infty$ and one can find that the entanglement entropy will grow linearly~\cite{Hartman:2013qma}.

Eq.~\eqref{timekappa} shows that relationship between $t_B$ and $\kappa_0$. Now let us consider the EoP between two regions $A$ and $B$. The entanglement of cross section is determined by $\kappa_0$ only, which can give us the relationship between $t_B$ and the EoP between two regions $A$ and $B$. For convenience, we will return to the coordinates $\{t,z,\vec{x}\}$.
Because of the symmetry, the entanglement wedge cross section $\Gamma(t_B)$is given by (see Fig.~\ref{FigAdS})
\begin{equation}\label{ewcG1}
  \Gamma(t_B):=\{x_1=x_1(s),z=z(s), -\infty<x_i<\infty,i=2,3,\cdots,d-1\}
\end{equation}
where $(x_1(s),z(s))$ satisfies the boundary conditions
\begin{equation}\label{boundcondt1}
  x_1(0)=0,z(0)=z_0\,
\end{equation}
with $\cos(d\kappa_0/2)^{2/d}=1/z_0$ and makes the area of cross section
\begin{equation}\label{areaewc}
  S_{\Gamma}:=V_{d-2}\int_0^\infty z^{1-d}\sqrt{|\dot{x}_1^2+\dot{z}^2/f|}\td s
\end{equation}
to be extremal. Here the dot means the derivative with respective to parameter $s$.


In order to obtain a well-proposed variational problem for integration~\eqref{areaewc}, we need to specify additional boundary conditions when $s\rightarrow\infty$. This can be achieved by imposing following suppositional boundary conditions at $s=l\rightarrow\infty$:
\begin{equation}\label{suppobound1}
  x_1(l)=l,z(l)=\tilde{z}_0, ~~~l\rightarrow\infty,
\end{equation}
where $\tilde{z}_0$ is an unknown parameter.
By using the boundary conditions~\eqref{suppobound1}, the extremal value of entanglement wedge cross section reads
\begin{equation}\label{Egammat}
  E(t_B):=E_P(A:B)=\frac{V_{d-1}\tilde{z}_0^{1-d}}4+\frac{V_{d-2}}4\int_0^\infty \left[z^{1-d}\sqrt{|\dot{x}_1^2+\dot{z}^2/f|}-\tilde{z}_0^{1-d}\dot{x}_1\right]\td s\,.
\end{equation}
Here $V_{d-1}:=V_{d-2}\int_0^\infty\td x_1$. As the ``Lagrangian'' in \eqref{areaewc} does not contain $x_1$ explicitly, the ``canonical momentum'' corresponding to $x_1$ is conserved. We can obtain following first order differential equation
\begin{equation}\label{firstorode1}
  p_0=z(s)^{1-d}\dot{x}_1(s)/\sqrt{|\dot{x}_1^2+\dot{z}^2/f|}
\end{equation}
with a constant $p_0$. By using the freedom of reparameterization, we can choose that $s=x_1$. Then the boundary conditions~\eqref{suppobound1} imply that $\dot{z}(\infty)=0$. Thus we obtain
\begin{equation}\label{valueofp0}
  p_0=\tilde{z}_0^{1-d}\,,
\end{equation}
and ~\eqref{firstorode1} leads to
\begin{equation}\label{firstorode2}
  x_1(z)=-p_0\int_{z_0}^{z}\frac{\td y}{\sqrt{f(y)(y^{2-2d}-p_0^2)}}=-\tilde{z}_0^{1-d}\int_{z_0}^{z}\frac{\td y}{\sqrt{f(y)(y^{2-2d}-\tilde{z}_0^{2-2d})}}\,.
\end{equation}
%
The boundary conditions~\eqref{suppobound1} require $x_1(\tilde{z}_0)\rightarrow\infty$, which implies
\begin{equation}\label{eqforp0ko}
  \left.\frac{\td}{\td y}\left[f(y)(\tilde{z}_0^{2-2d}-y^{2-2d})\right]\right|_{y=\tilde{z}_0}=0\,.
\end{equation}
This shows that $\tilde{z}_0=z_h=1$. Then we obtain
\begin{equation}\label{Egammat2}
\begin{split}
  E(t_B)&=\frac{V_{d-1}}4+\frac{V_{d-2}}4\int_0^\infty \left[z^{1-d}\sqrt{\dot{x}_1^2+\dot{z}^2/f}-\dot{x}_1\right]\td s\\
  &=\frac{V_{d-1}}4+\frac{V_{d-2}}4\int_1^{z_0} \left[z^{1-d}\sqrt{x_1'^2+1/f}+\frac{\td x_1}{\td z}\right]\td z
  \end{split}\,.
\end{equation}
and  Eq.~\eqref{firstorode2} gives
\begin{equation}\label{xprime}
  \frac{\td x_1}{\td z}=-\frac{1}{\sqrt{f(z)(z^{2-2d}-1)}}.
\end{equation}
Combining  above two results, we obtain
\begin{equation}\label{Egammat3}
\begin{split}
  E(t_B)=\frac{V_{d-1}}4+\frac{V_{d-2}}4\int_1^{z_0}\sqrt{\frac{z^{2-2d}-1}{1-z^d}}\td z
  \end{split}\,.
\end{equation}
Here $t_B$ and $z_0$ are connected by Eq.~\eqref{timekappa} with $\cos(d\kappa_0/2)^{2/d}=1/z_0$.

For the case that $d=2$, the relationship between $t_B$ and $\kappa_0$ can be computed analytically, which yields~\cite{Hartman:2013qma}
\begin{equation}\label{realkappa0tB}
  \sinh t_B=\tan\kappa_0,~~z_0=\frac1{\cos\kappa_0}=\cosh t_B~~~\kappa_0\in[0,\pi/2)\,.
\end{equation}
The Eqs.~\eqref{Egammat} and \eqref{ETFDs1} turn to
\begin{equation}\label{Egammatd2}
  E(t_B)
  =\frac{V_{1}}4+\frac14\ln\left(\cosh t_B\right),~~S_{AB}(t_B)=\frac12\ln\left(2\cosh t_B\right)+2S_{\text{div}}\,.
\end{equation}
We see that, up to constant factors, the holographic entanglement entropy between $A\cup B$ and $\overline{A\cup B}$ and the holographic EoP between $A$ and $B$  have the similar time dependent behavior,
\begin{equation}\label{growthEopS1}
  \frac{\td}{\td t_B}E(t_B)=\frac12\frac{\td}{\td t_B}S_{AB}(t_B)\,.
\end{equation}

\begin{figure}
  \centering
  \includegraphics[width=.48\textwidth]{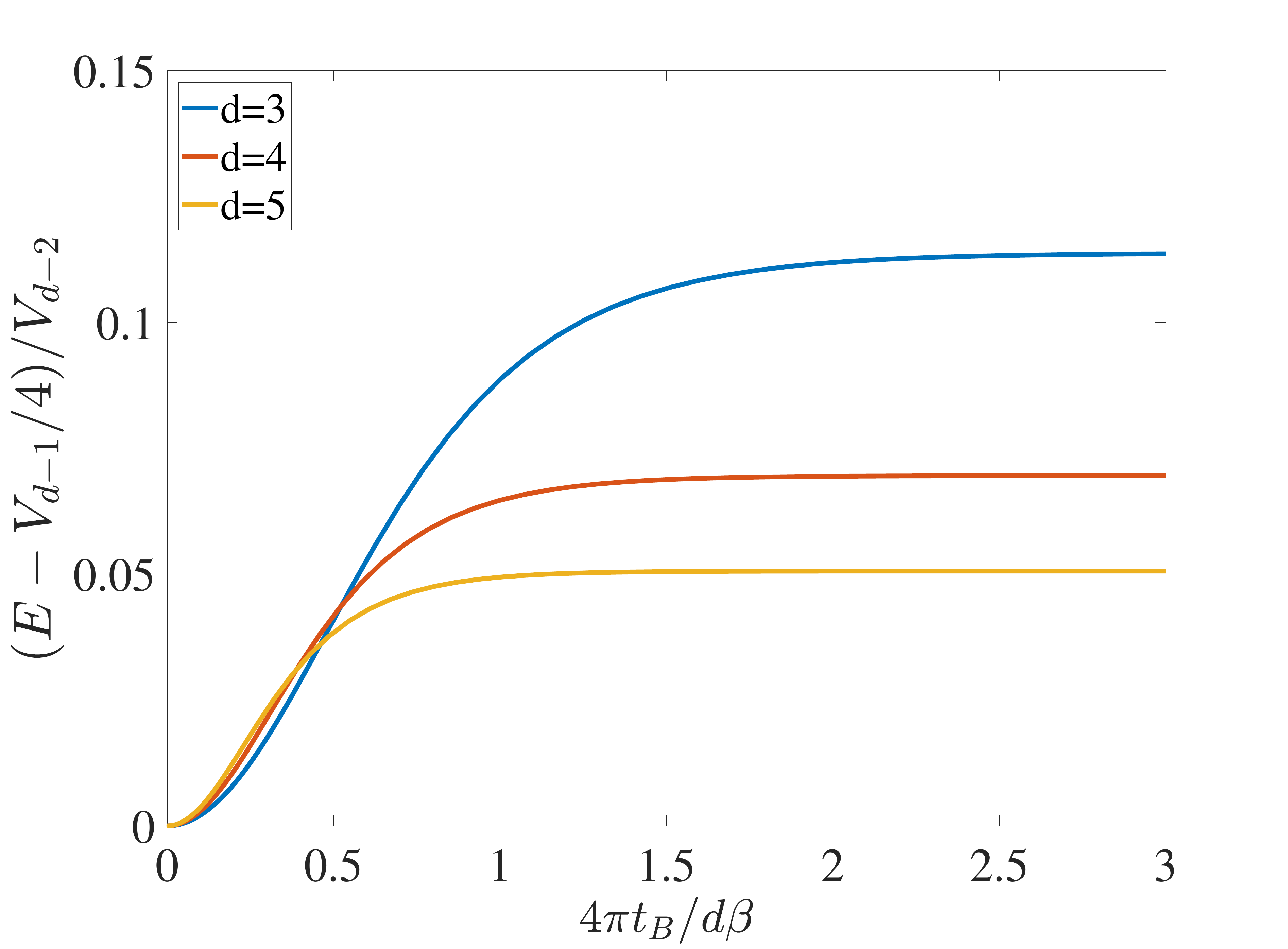}
  \includegraphics[width=.48\textwidth]{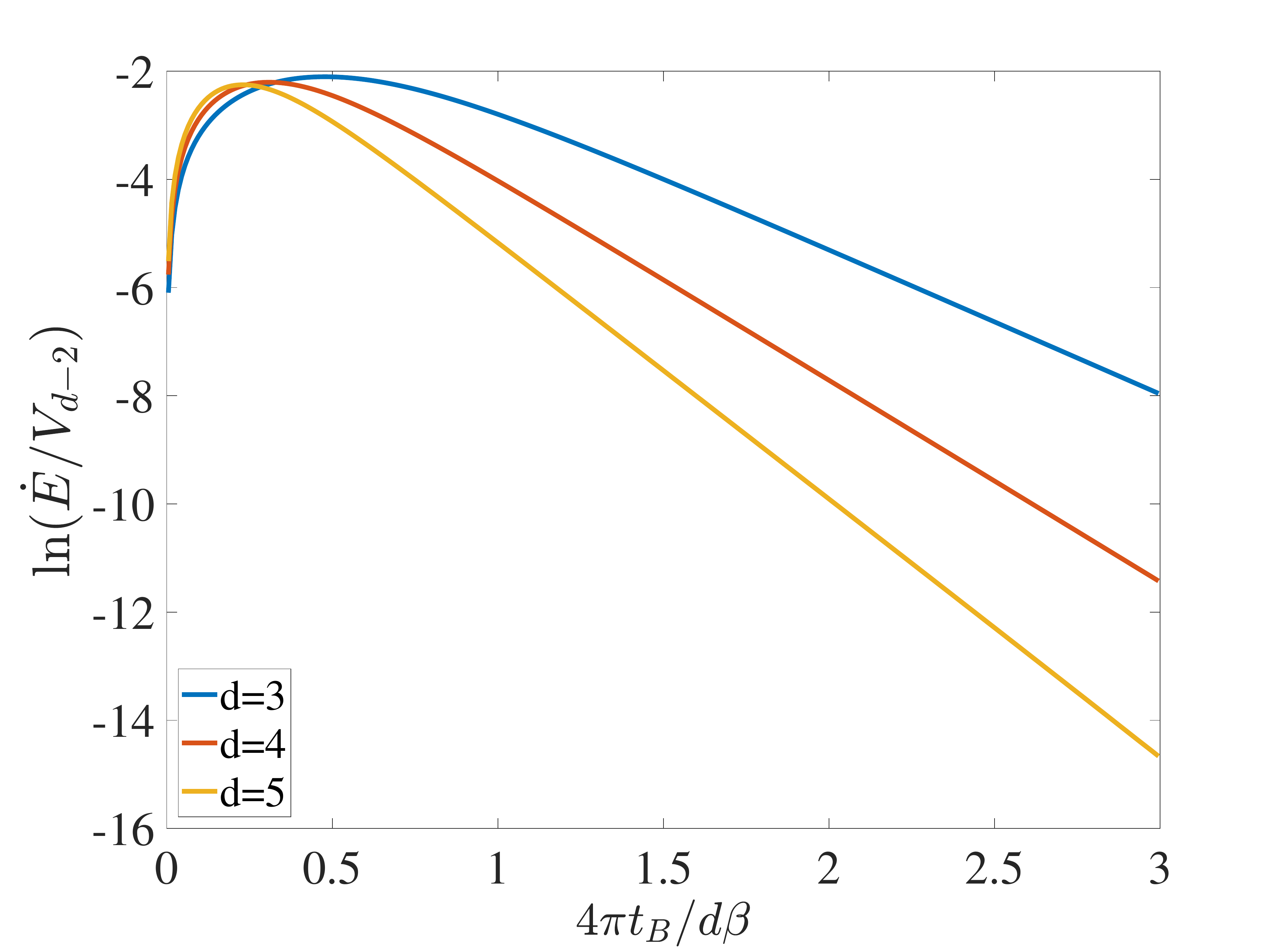}
  \caption{Relationship between entanglement of purification and time $t_B$ when $d=3,4,5$. Here we set $z_h=1$ and $\Delta E(t_B)=E(t_B)-E(0)$. } \label{plotTFD1}
\end{figure}

When $d>2$, there is an essential difference between the holographic entanglement entropy and the holographic EoP. From  ~\eqref{timekappa}, we can find~\cite{Hartman:2013qma}
\begin{equation}\label{regionk0}
  \kappa_0\in\left[0,\frac2d\arcsin\sqrt{d/(2d-2)}\right).
\end{equation}
The corresponding $z_0$ then satisfies
\begin{equation}
1\leq z_0<\frac{2d-2}{\sqrt{(d-2)(3d-2)}}\,.
\end{equation}
In the case that $d=2$, $z_0$ will approach to infinity when $t_B\rightarrow\infty$ and so $E(t_B)-V_1/4$ can approach to infinity. However, in the case $d>2$, $z_0$ will approach to $\frac{2d-2}{\sqrt{(d-2)(3d-2)}}$ when $t_B\rightarrow\infty$. Then ~\eqref{Egammat3} implies $E(t_B)-V_{d-1}\tilde{z}_0^{d-1}/4$ will approach to a finite value. There is no compact the analytical solution for both holographic entanglement entropy and holographic EoP in higher dimension. The time evolutional behaviors of $E(t_B)$ and $\dot{E}:=\td E(t_B)/\td t_B$ are shown in Fig.~\ref{plotTFD1}.
It has been show that the holographic entanglement entropy between $A\cup B$ and $\overline{A\cup B}$ will grow linearly when $t_B\gg\beta$~\cite{Hartman:2013qma}. From the numerical results shown in the left panel of Fig.~\ref{plotTFD1}, we can see that the holographic EoP between $A$ and $B$  always grows. However, the speed of growth decays exponentially,
\begin{equation}\label{decay1}
  \dot{E}:=\td E(t_B)/\td t_B\propto\exp(-\alpha_dt_B/\beta)\,.
\end{equation}
Here $\alpha_d$ is a constant which depends on the dimension $d$.

\subsection{Finite size case}
Now let us consider two finite subregions which are in the two boundaries of the spacetime respectively.  The subregions $A$ and $B$ are given by
$$A=\{t_L=t_B, 0<x_1<l, -\infty<x_i<\infty,i=2,3,\cdots,d-1\}$$
and
$$B:=\{t_R=t_B, 0<x_1<l,, -\infty<x_i<\infty,i=2,3,\cdots,d-1\}\,.$$
\begin{figure}
  \centering
  \includegraphics[width=.46\textwidth]{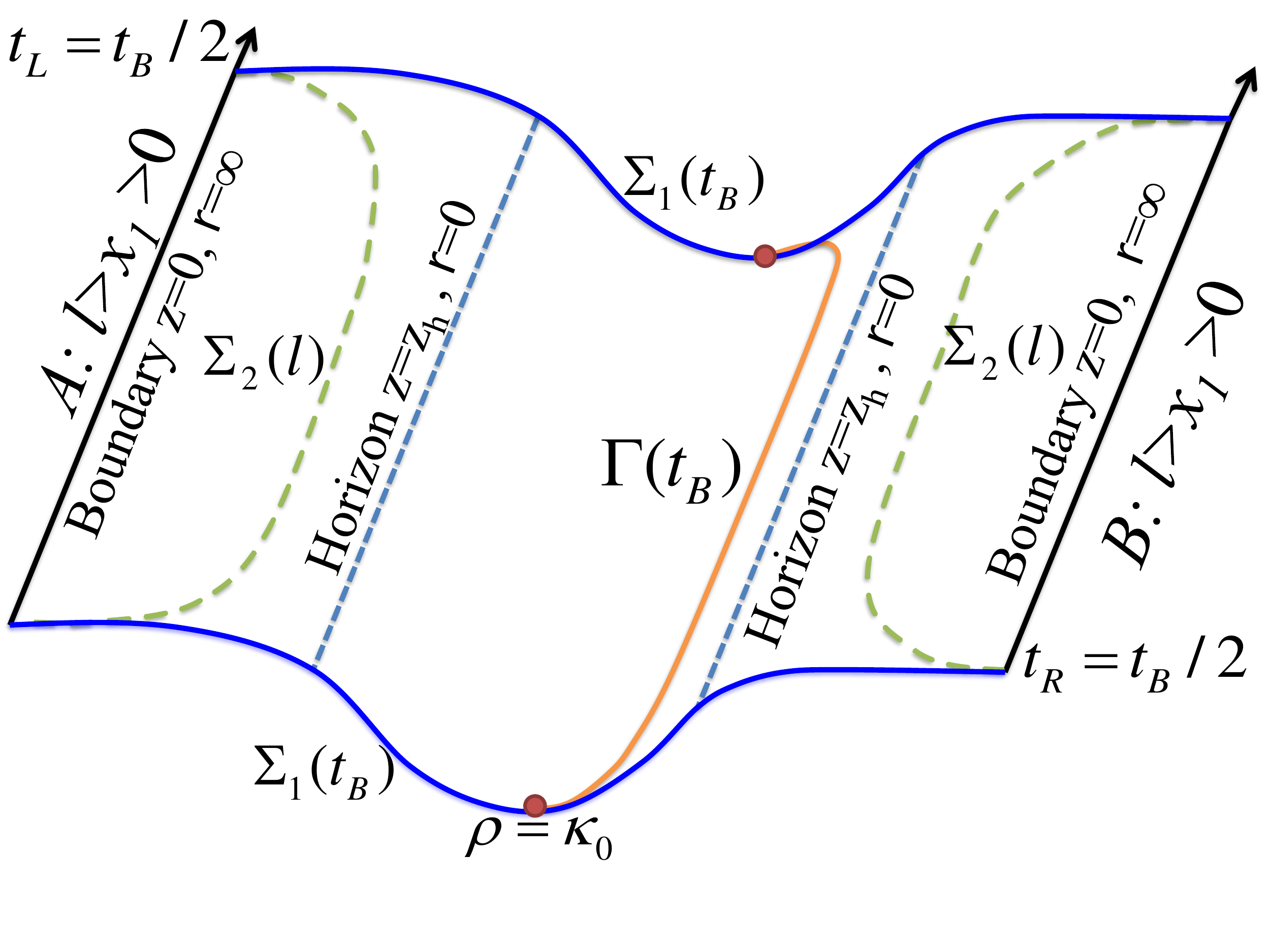}
  \includegraphics[width=.46\textwidth]{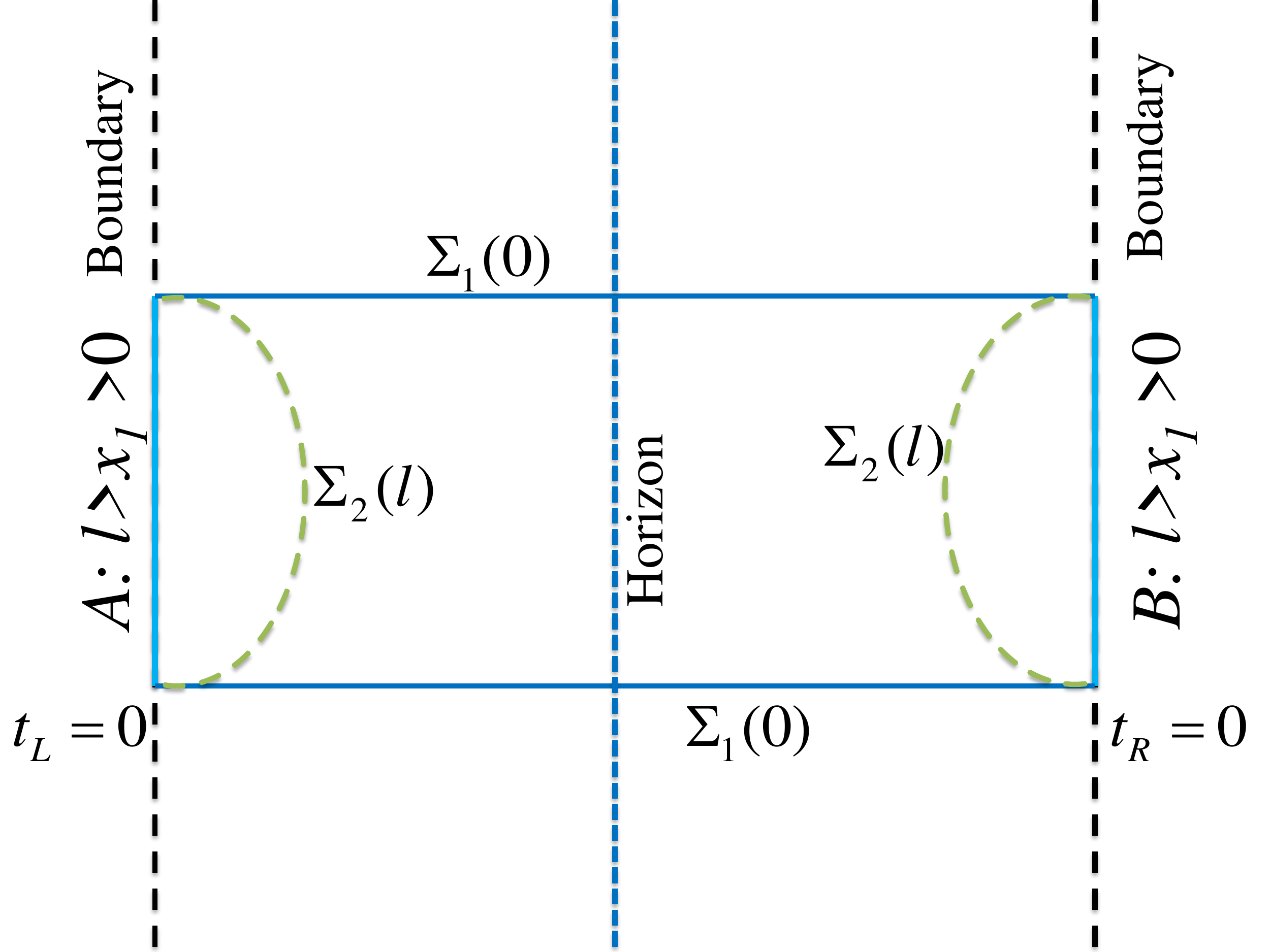}
  \caption{The case that two finite subregions are in the different boundaries of black brane. The left panel is the case $t_L=t_R=t_B>0$ and the right panel is the case that $t_L=t_R=0$. } \label{figTFD2}
\end{figure}
The induced density matrix of $A\cup B$ is also time-dependent,
\begin{equation}\label{inducedAB1}
  \rho_{AB}=\Tr_{L,0<x_1<l}\Tr_{R,0<x_1<l}(|\Phi(t_B)\rangle\langle\Phi(t_B)|)\,.
\end{equation}
Thus, the entanglement of purification between $A$ and $B$ depends on time $t_B$ and the size $l$. Similar to the infinitely wide case, we can also compute the entanglement entropy for the union $A\cup B$ by ~\eqref{defSABTFD}, which is dependent on the width $l$ and boundary time $t_B$.

\begin{figure}
  \centering
  \includegraphics[width=.49\textwidth]{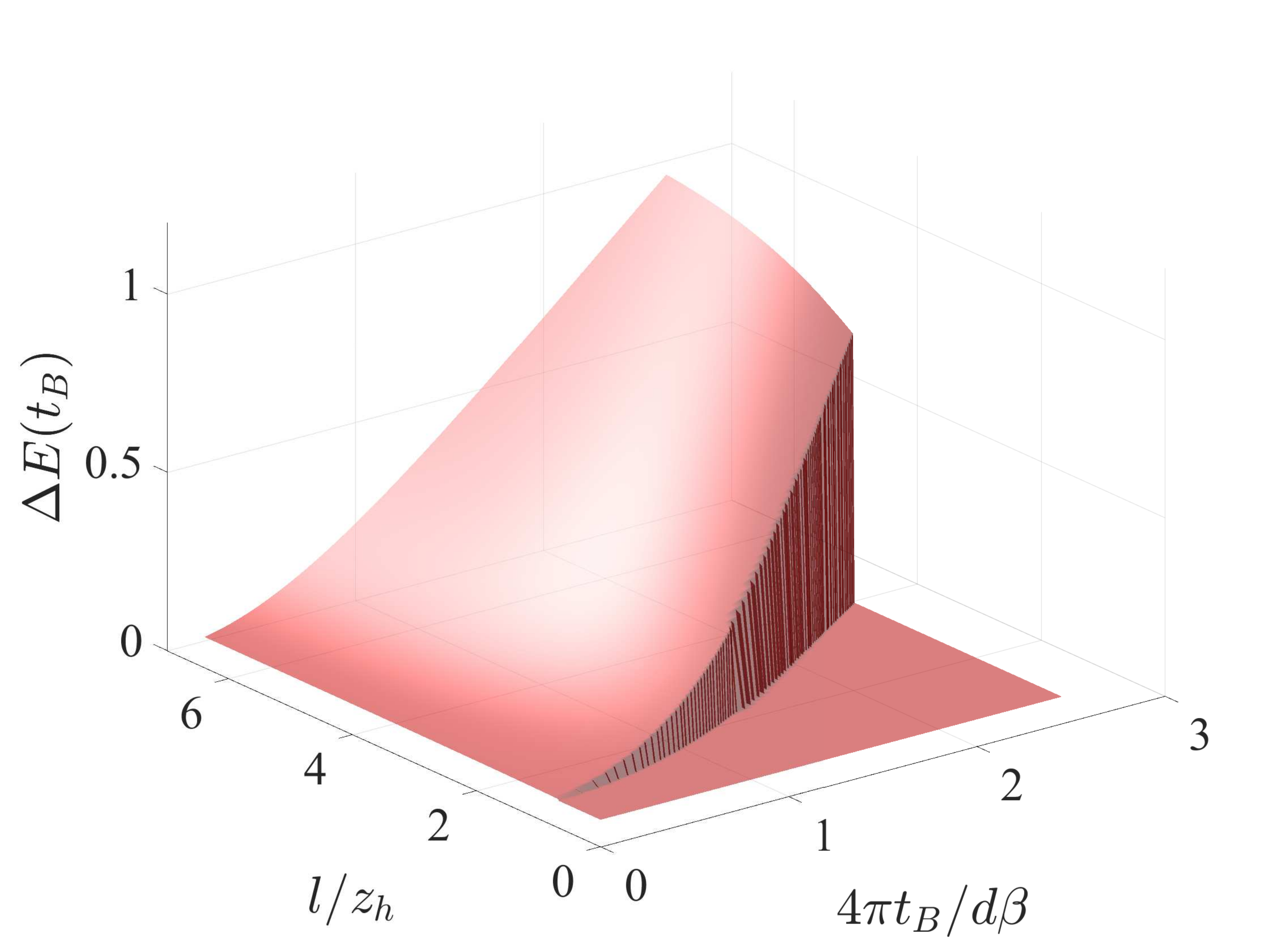}
  \includegraphics[width=.49\textwidth]{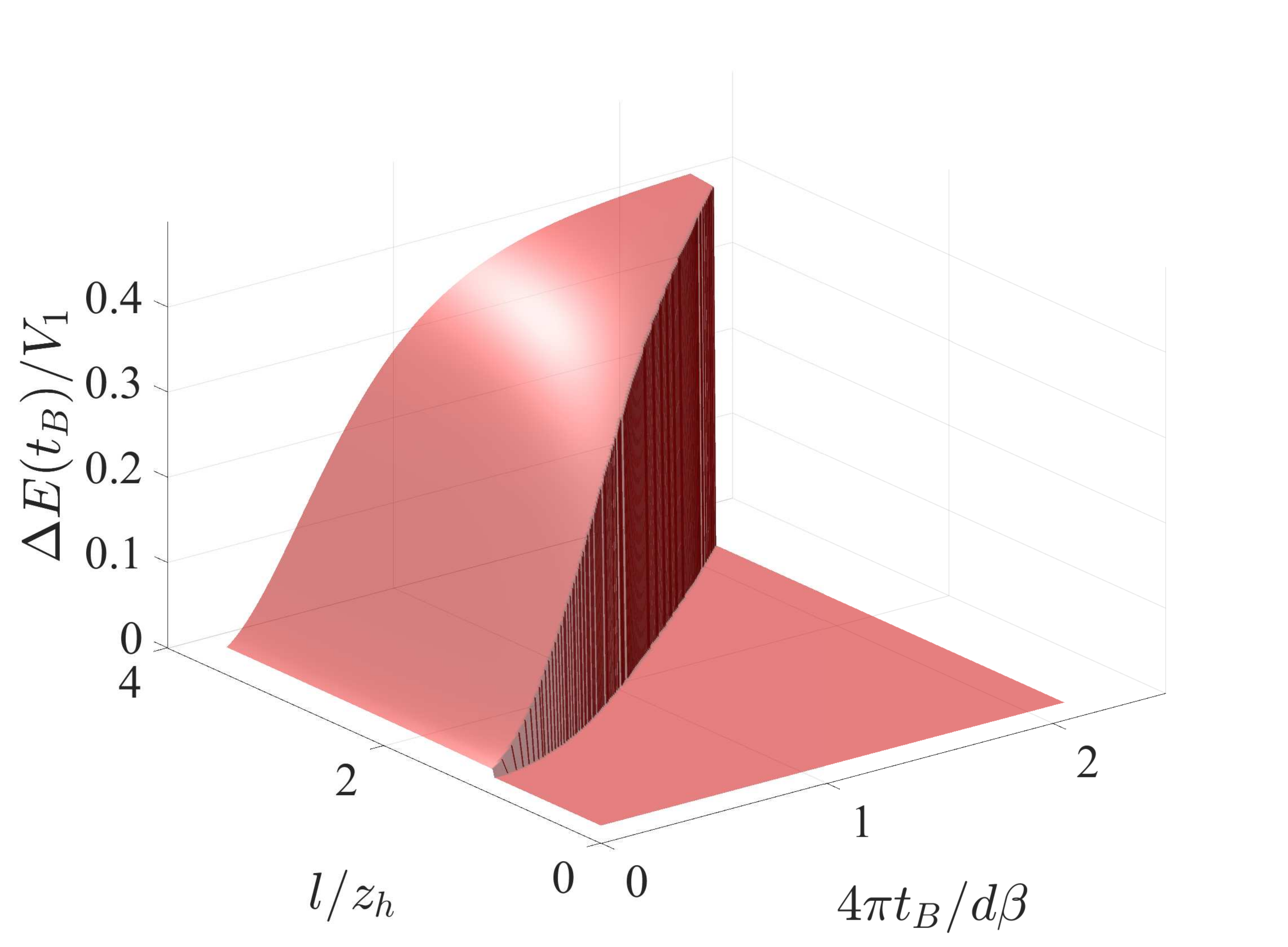}
  \caption{Relationship between holographic EoP for different $l$ and time $t_B$ when $d=2$ (left panel) and $d=3$ (right panel). $\Delta E(t_B)=E(t_B)-E(0)$. The cases $d>3$ are similar to the right panel.} \label{eopld}
\end{figure}

From the holographic viewpoint, we see that there are two possibilities as shown in the Fig.~\ref{figTFD2}. The first one is the case that the extremal surfaces connecting $\partial A$ and $\partial B$ are $\Sigma_2(l)$, which are two disconnected extremal surfaces at with fixed $t$ and will give zero entanglement wedge cross section. The second one is the case that the extremal surfaces connecting $\partial A$ and $\partial B$ are $\Sigma_1(t_B)$, which connects two subregions $A$ and $B$ and has a nonzero entanglement wedge cross section $\Gamma(t_B)$. The entanglement wedge is connected only if $\Sigma_1(t_B)<\Sigma_2(l)$.
We see that the nonzero initial holographic EoP $E(l,t_B=0)$ can appear when the area of $\Sigma_1(0)$ is smaller than the area of $\Sigma_2(l)$. As the area of $\Sigma_2(l)$ is constant and zero if $l\rightarrow0$ but the  area of $\Sigma_1(t_B)$ is nonzero and  increases monotonously with $t_B$ ~\cite{Hartman:2013qma}, we can conclude that there is a critical length $l_c$ and a critical time $t_c(l)$ for $l>l_c$ such that
\begin{equation}\label{criticalls1}
  E(t_B)=\left\{
  \begin{split}
  0&,~~l\leq l_c,~\text{or}~t_B>t_c(l)\\
  \frac14\text{Area}(\Gamma(t_B))&,~~l>l_c~\text{and}~t_B<t_c(l)\,.
  \end{split}
  \right.
\end{equation}
Here $\Gamma(t_B)$ is the extremal surface corresponding to the entanglement wedge cross section which connects the two segments of $\Sigma_1(t_B)$. Similar to the last subsection, $\Gamma(t_B)$ can be determined by finding extremal value of following integration
\begin{equation}
\int_0^l z^{1-d}\sqrt{|\dot{x}_1^2+\dot{z}^2/f|}\td s
\end{equation}
with the boundary conditions
\begin{equation}
x_1(0)=0,~z(0)=z_0,~x_1(l)=l~,z(l)=z_0\,.
\end{equation}
Using the similar steps in last subsection, we can obtain the expression for the area of extremal surface
\begin{equation}\label{areagamma1}
  \text{Area}(\Gamma(t_B))=V_{d-2}l\tilde{z}_0^{1-d}+2V_{d-2}\int_{\tilde{z}_0}^{z_0}\frac{\tilde{z}^{1-d}_0-z^{2-2d}}{\sqrt{(1-z^d)(z^{2-2d}-\tilde{z}^{2-2d}_0)}}\td z,
\end{equation}
where $\tilde{z}_0$ is determined by $l$ according to following differential equation
\begin{equation}\label{newz0s1}
  \tilde{z}_0^{1-d}=z(x_1)^{1-d}/\sqrt{1+z'^2/f},~~~\left.\frac{\td z}{\td x_1}\right|_{x_1=l/2}=0\,.
\end{equation}
Thus, in the case that $l>l_c$ and $t_B<t_c(l)$, the evolution of EoP is given by  ~\eqref{areagamma1} and \eqref{newz0s1}. The EoP will first increase with respective to $t$ and turn to zero suddenly when $t>t_c(l)$. The values of $E(t_B)$ for different $l$ and $t_B$ are shown in the Fig.~\ref{eopld}.

The values of $l_c$ and $t_c(l)$ are determined as follows. Taking the time slices of $t_L=t_R=t_B=0$, one can find that the area of $\Sigma_1(0)$ is given by
\begin{equation}\label{defSigam1}
  S_{AB}(0):=V_{d-2}\int_{\epsilon}^{z_h}\frac{\td z}{z^{d-1}\sqrt{f(z)}}=\left\{
  \begin{split}
  &4S_{\text{div}},~~d=2\\
  &4S_{\text{div}}-\frac{V_{d-2}}{(d-2)}\left(\frac{\pi}{d\beta}\right)^{d-2}\frac{\sqrt{\pi}\Gamma(2/d)}{\Gamma(2/d-1/2)},~~d>2\,.
  \end{split}
  \right.
\end{equation}
Here $S_{\text{div}}$ is given by ~\eqref{defUVS} and $\beta=4\pi z_h/d$. The area of $\Sigma_2(l)$ is given by
\begin{equation}\label{minimL}
  S_2(l)=2S(l)
\end{equation}
and $S(l)$ is defined in  ~\eqref{eq:OneSideSz}. Then the value of $l_c$ is determined by
\begin{equation}\label{valuelc}
  S_2(l_c)=S_{AB}(0)\,.
\end{equation}
When $t_L=t_R=t_B\neq0$, the area of $S_{AB}(t_B)$ is given by  ~\eqref{ETFDs1}, which increases monotonously  with  $t_B$. The value of $t_c$ is then determined by
\begin{equation}\label{valuetc}
  S_2(l)=S_{AB}(t_c)\,.
\end{equation}

 When $d=2$, we can obtain analytical results for $l_c$ and $t_c$.  $S_{AB}(t_B)$ is the same as Eq.~\eqref{Egammatd2} and $S_2(l)$ is given by
\begin{equation}\label{BTZSS12}
  S_2(l)=4S_{\text{div}}+\ln\sinh\frac{\pi l}{\beta}\,.
\end{equation}
Solving \eqref{valuelc} and \eqref{valuetc}, we get
\begin{equation}
l_{c}=\frac{\beta}{\pi}\ln(\sqrt{2}+1),~~~ t_{c}(l)=\frac{\beta}{2\pi}\text{arccosh}\left(\sinh\frac{\pi l}{\beta}\right).
\end{equation}
In the limit $l\gg\beta$, we have $t_c(l)\approx l/2$.
The entanglement of purification between $A$ and $B$ is shown in the left panel of Fig.~\ref{eopld}.

\begin{figure}
  \centering
  \includegraphics[width=.45\textwidth]{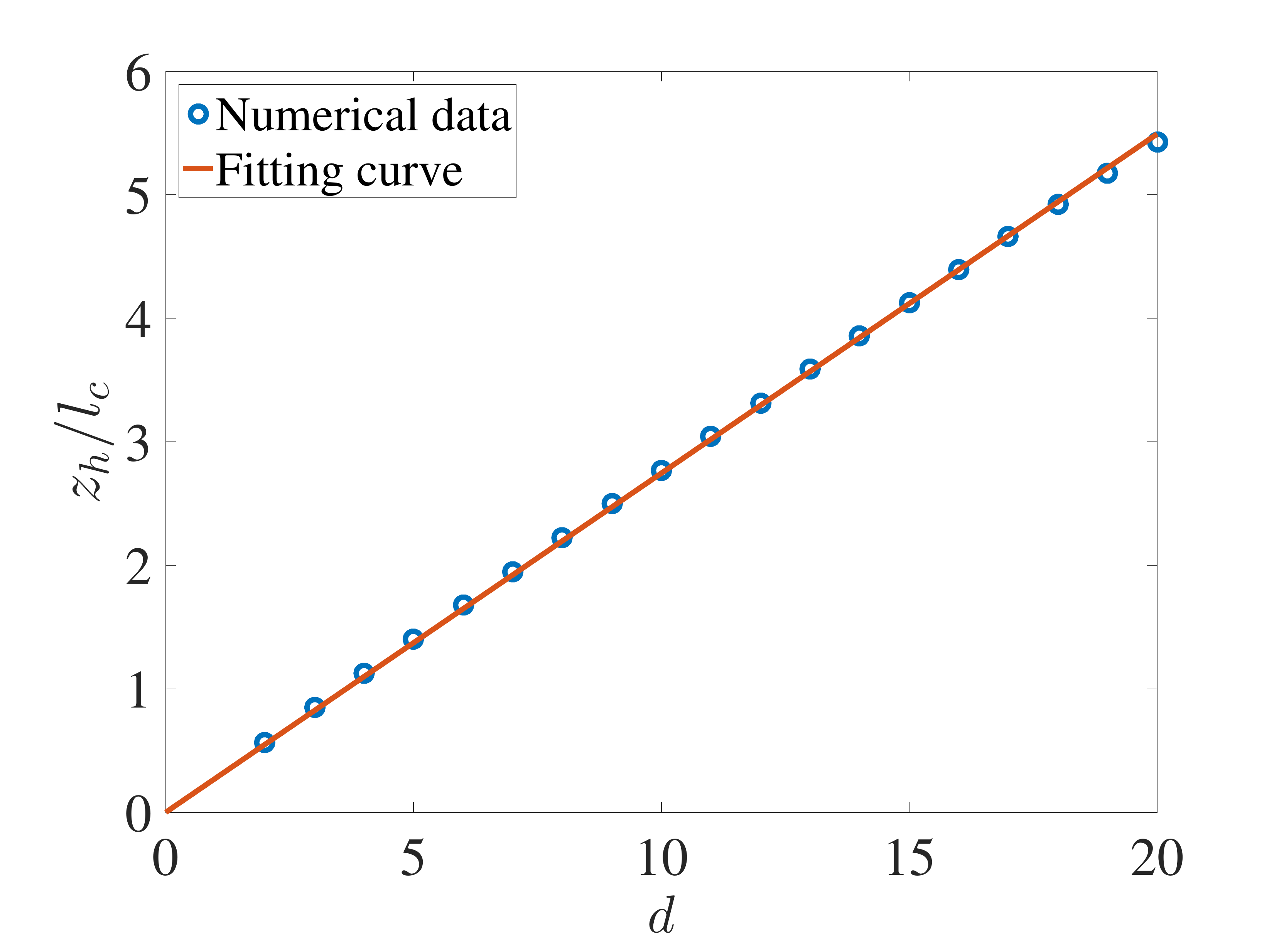}
  \includegraphics[width=.45\textwidth]{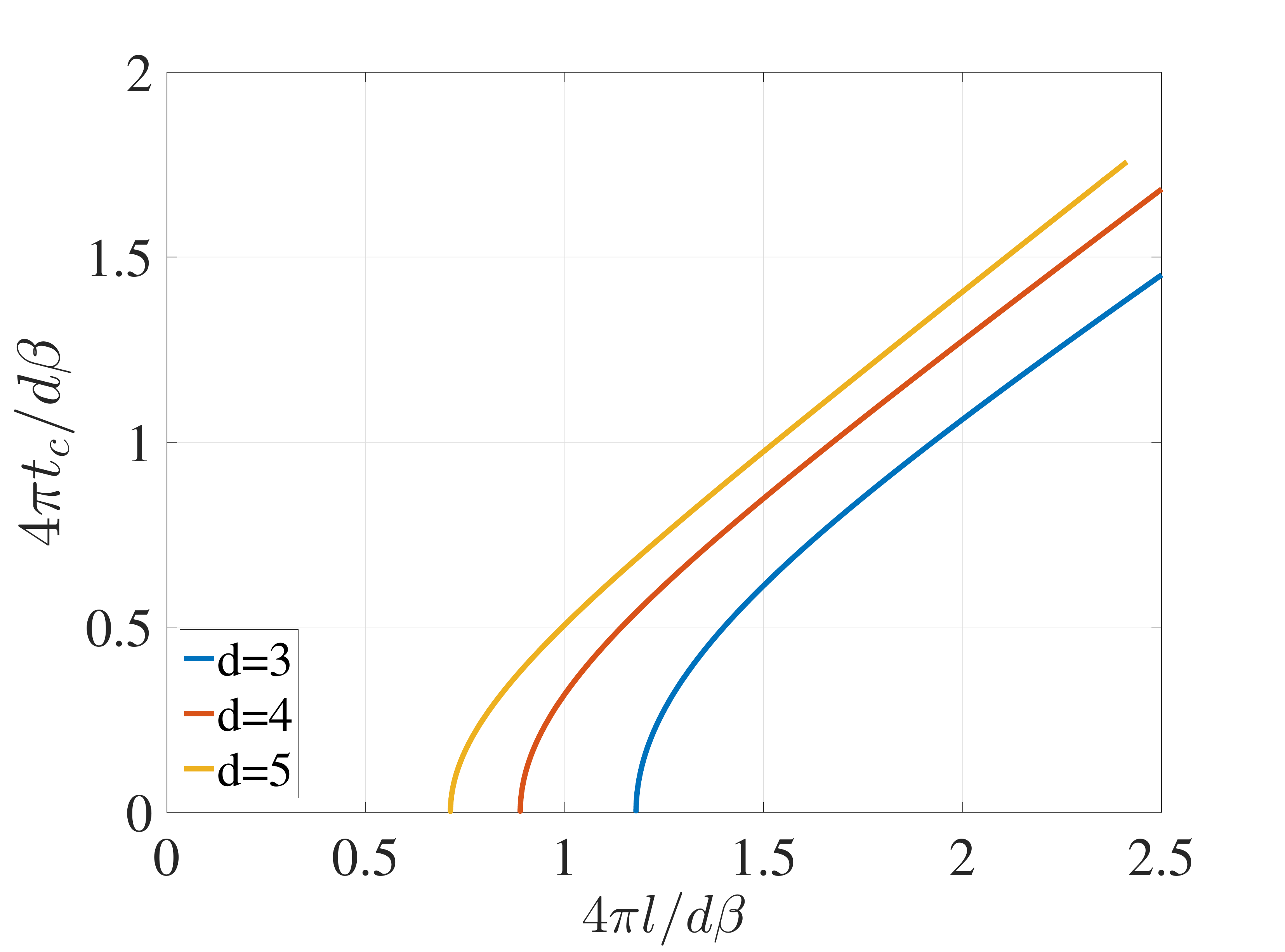}
  \caption{Left panel: the values of $l_c$ at different dimensions. In the region of $d=2,3,4,\cdots,20$, we find that the $z_hl_c^{-1}\approx0.27d$. Right panel: relationship between $t_c$ and $l$ when $d=3,4,5$. } \label{plottcl}
\end{figure}

 When $d>2$, there is no compact analytical results for $l_c$ and $t_c$.
The values of $l_c$ and $t_c(l)$ can be obtained numerically by using ~\eqref{eq:OneSideSz}, \eqref{eq:OneSideLz}, \eqref{timekappa} and \eqref{ETFDs1}. The results are shown in Fig.~\ref{plottcl}.
%
%
From the right panel, we see that $t_c(l)$ increases monotonously with $l$. From the left panel, we see that $l_c$ can be approximated well by
\begin{equation}
z_h/l_c\approx0.27d\Rightarrow l_c\approx\frac{\beta}{\pi}\ln(\sqrt{2}+1)\,.
\end{equation}
This is similar to the case of BTZ black brane.
When the size $l$ of $A$ and $B$ are much larger than $\beta$ and $t_B\gg\beta$, $S_2(l)$ and $S_{AB}(t_B)$ depend on $l$ and $t_B$ linearly, respectively.
\begin{equation}\label{asyptlargel}
   S_2(l)\approx \frac12lV_{d-2}+4S_{\text{div}},~~~S_{AB}(t_B)\approx v_d V_{d-2}t_B+4S_{\text{div}}.
\end{equation}
Here $v_2=1$ and $v_d=\sqrt{d}(d-2)^{\frac12-\frac1d}/[2(d-1)]^{1-\frac1d}\in(1/2,1)$ for $d>2$. Thus, we can see that,
\begin{equation}\label{asyptlarge2}
  t_c(l)= \frac{l}{2v_d},~~~l\gg\beta\,.
\end{equation}
In the case of $d\gg1$, we have $t_c\approx l$. This agrees with the numerical results shown in the right panel of Fig.~\ref{plottcl}.

\section{Evolution of EoP after a thermal quench}\label{quenchEoP}
\begin{figure}
\begin{centering}
\includegraphics[scale=0.35]{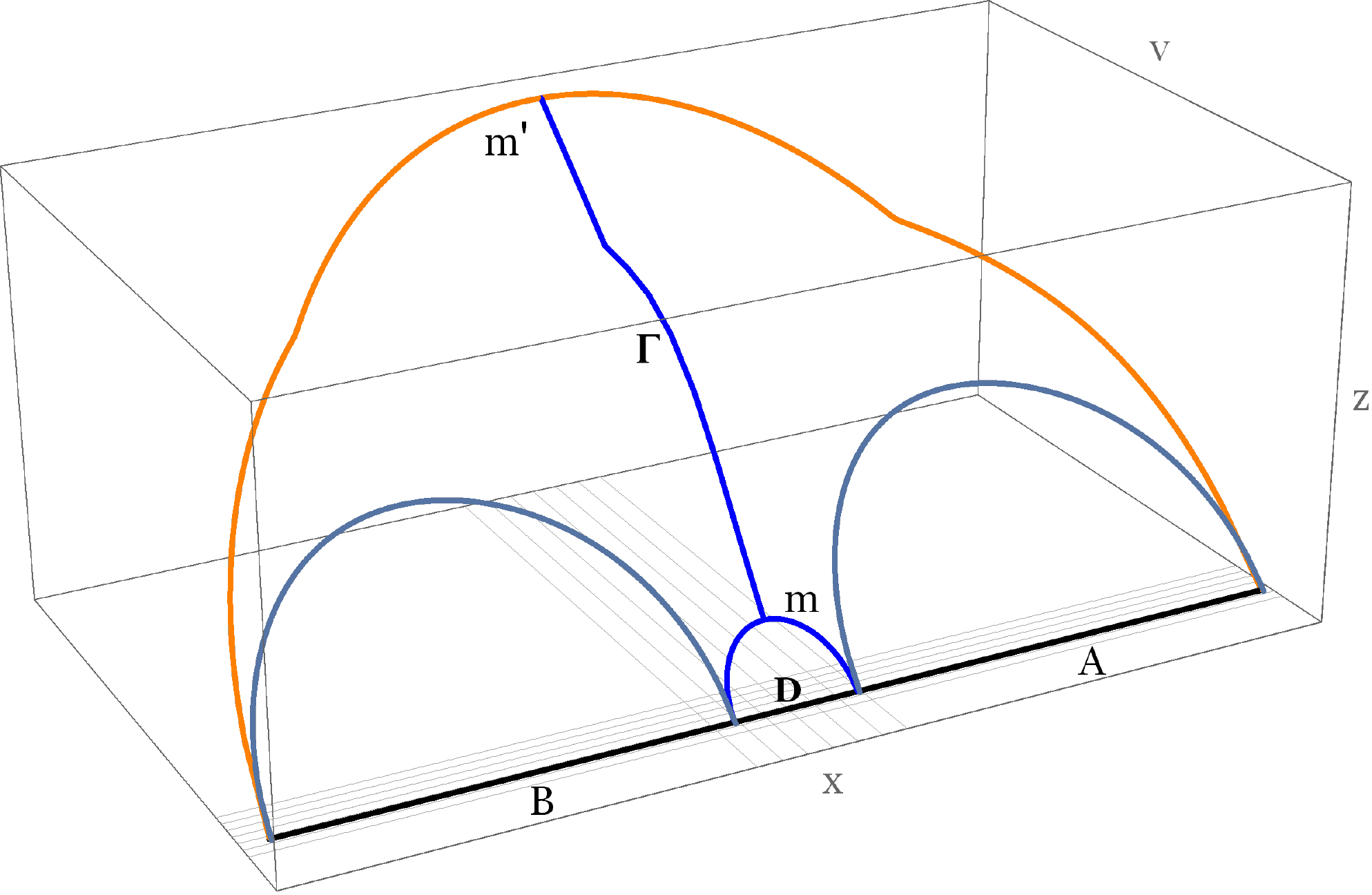}
\par\end{centering}
\caption{\label{fig:HRT}The Hubeny-Rangamani-Takayanagi surfaces for finite
strips in Vaidya-AdS spacetime. $m$ is the turning point of the HRT
surface corresponding to strip with width $D$ and $m'$ to $2l+D$.
The area of the extremal surface $\Gamma$ connecting $m$ and $m'$ is proportional
to the EoP.}
\end{figure}

In this section, we consider the evolution of EoP in CFT after
a thermal quench. This process can be described holographically by
the Vaidya-AdS metric which reads
\begin{align}
ds^{2}= & \frac{1}{z^{2}}\left[-f(v,z)dv^{2}-2dzdv+dx^{2}+\sum_{i=1}^{d-2}dy_{i}^{2}\right],\label{eq:VaidyaMetric}\\
f(v,z)= & 1-m(v)z^{d}.\nonumber
\end{align}
Here the AdS space radius is rescaled to $1$. The mass function
we take is
\begin{equation}
m(v)=\frac{M}{2}\left(1+\tanh\frac{v}{v_{0}}\right),\label{eq:Massfunction}
\end{equation}
 where $v_{0}$ characterizes the quench speed. We fix $v_{0}=0.01$ without loss of generality and set $M=1$ in this section for simplicity.
 When $v\to-\infty$, the spacetime
is pure AdS. When $v\to\infty$, the spacetime becomes a planar
SAdS black brane.
We consider two finite strips $A$ and $B$ with the same width $l$ on one side.
The separation is $D$. See Fig. \ref{fig:HRT} for the configuration. Due to the translation symmetry of the metric, the entanglement wedge of cross section $\Gamma$ lies in the $(z,v)$ plane as shown in Fig.~\ref{fig:HRT}.

\begin{figure}
\begin{centering}
\begin{tabular}{cc}
\includegraphics[scale=0.45]{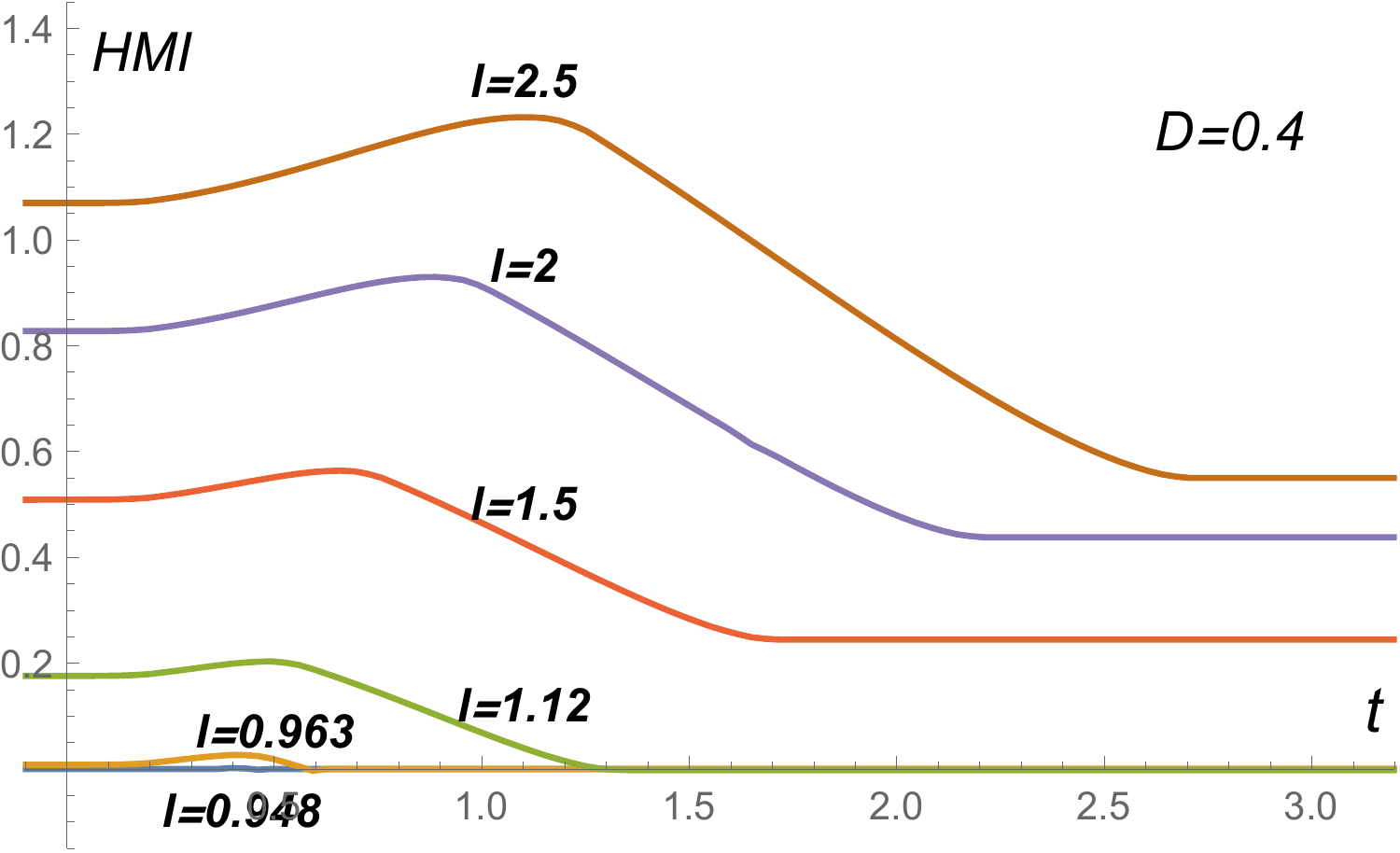} & \includegraphics[scale=0.43]{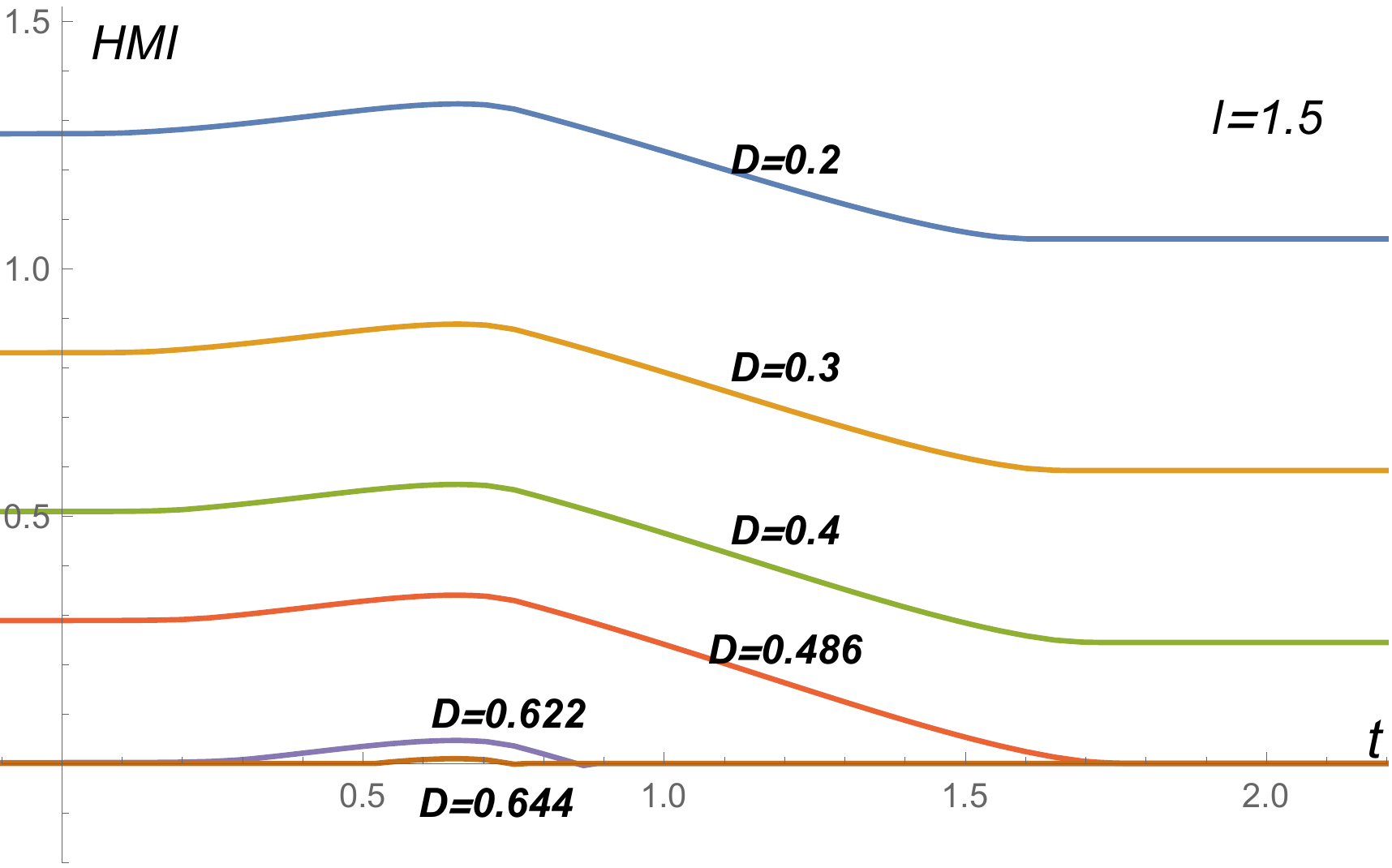}\tabularnewline
\end{tabular}
\par\end{centering}
\caption{\label{fig:VaidyaHMItime}The evolution of HMI (in unit of $4/V_{d-2}$) for given $l$ and separation
$D$ when $d=2$. In left panel, we fix the separation $D=0.4$. In right panel, we fix the width $l=1.5$.}
\end{figure}

\begin{figure}
\begin{centering}
\begin{tabular}{cc}
\includegraphics[scale=0.45]{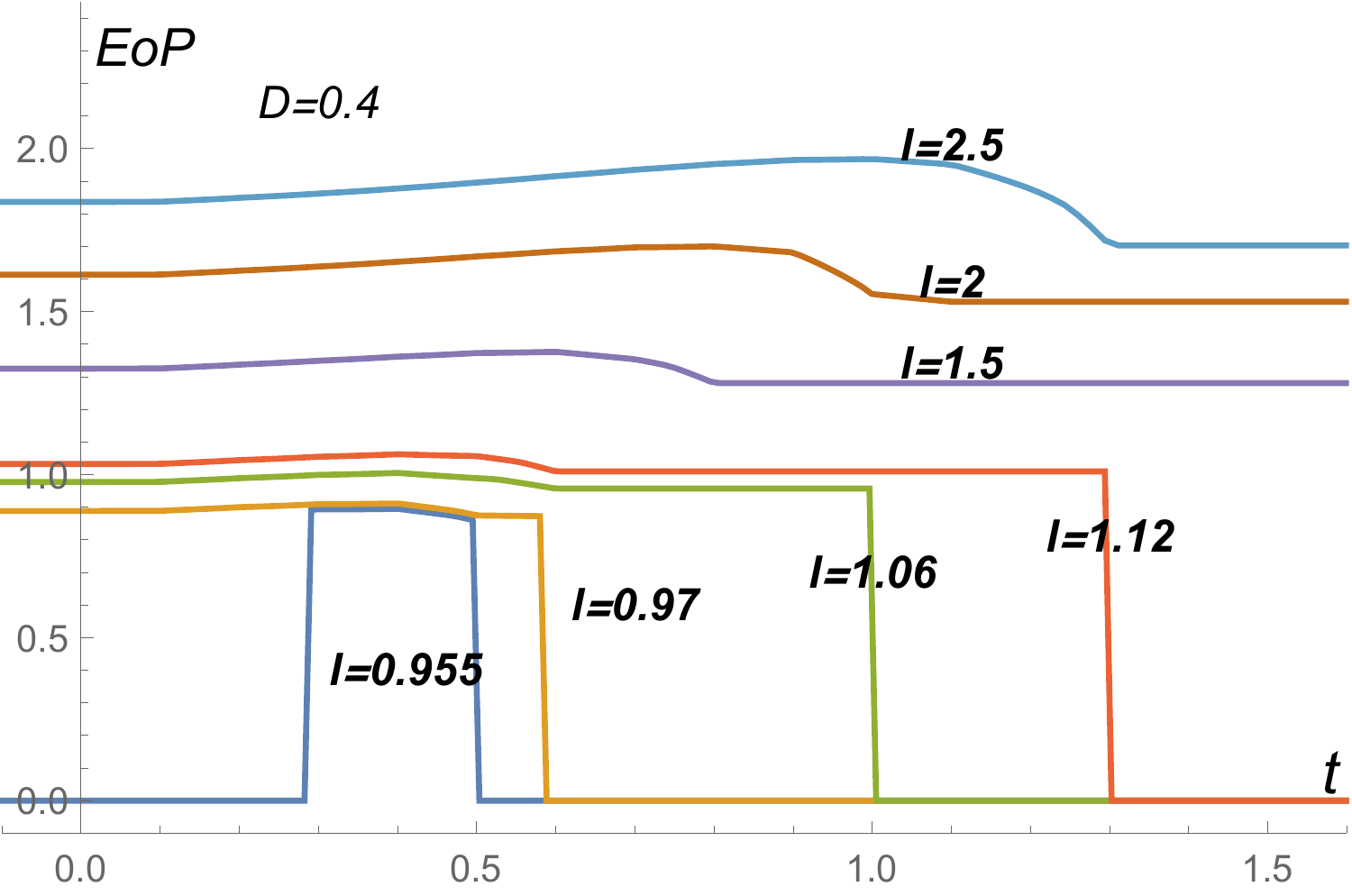} & \includegraphics[scale=0.45]{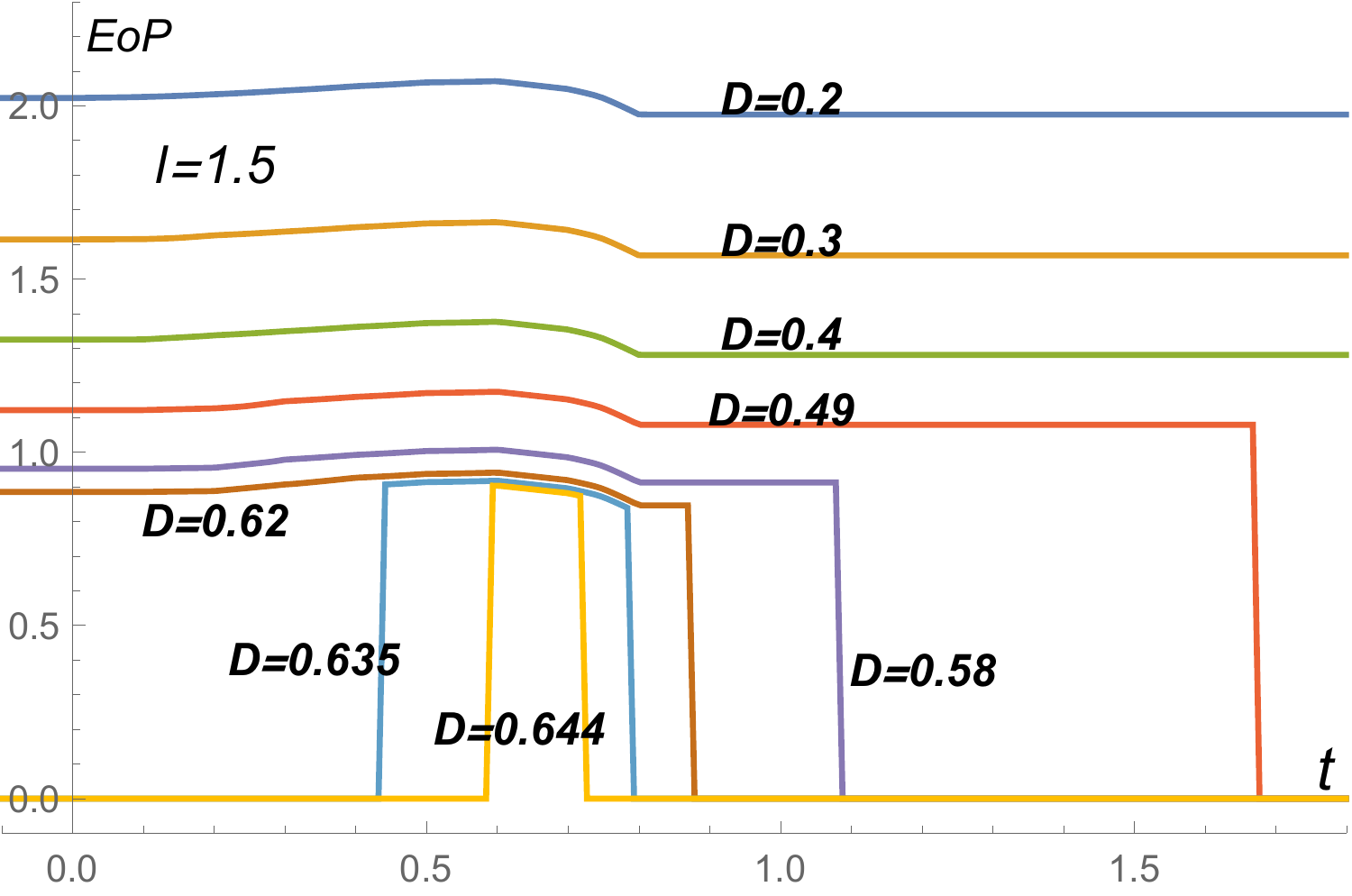}\tabularnewline
\end{tabular}
\par\end{centering}
\caption{\label{fig:VaidyaEoPtime}The evolution of holographic EoP (in unit of $4/V_{d-2}$) for given $l$ and separation
$D$ when $d=2$. We fix $D=0.4$  in the left panel and $l=1.5$ in the right panel, respectively.}
\end{figure}

The EoP exists only when the MI
\begin{equation}
I(l,D,t)=2S_{l}(t)-S_{D}(t)-S_{2l+D}(t)>0.\label{eq:VaidyaEoPCondition}
\end{equation}
 Here $S_{w}(t)$ is the entanglement entropy corresponding to a strip
with width $w$ at the boundary time $t$. It can be calculated holographically
by the area of the corresponding HRT surface.
\begin{equation}
S_{w}(t)=\frac{2V_{d-2}}{4 }\int_{\delta}^{z_{0}(w,t)}\frac{dz}{z^{d-1}}\frac{1}{\sqrt{f(v,z)\left(1-\frac{z^{2d-2}}{z_{0}^{2d-2}}\right)}}.
\end{equation}
 Here $z_{0}(w,t)$ is the turning point of the corresponding HRT surface at boundary time $t$. The HRT surface can be worked out numerically, see \cite{Chen:2018mcc} for example. We show the evolution of HMI\footnote{For more works on the evolution of mutual information, see \cite{Allais:2011ys,Ziogas:2015aja}.} when $d=2$ in Fig. \ref{fig:VaidyaHMItime}. In the left panel, we fix the separation $D=0.4$ and let $l$ run. In the right panel, we fix the width $l=1.5$ and let $D$ run. In both panels, the HMI grows with time at first and then decreases to a value that is smaller than the initial value. The equilibrium time is about $l+D/2$.
 In the left panel, when $l>1.12$, the HMI is always greater than zero in the evolution process. Namely, there is always holographic EoP between strips $A$ and $B$. When $1.12>l>0.963$, the HMI is greater than zero at first and then decreases to zero, i.e., there is holographic EoP at first but it vanishes later. When $0.963>l>0.948$, the quantity $I(l,D,t)$ is negative at first, then it grows to be positive but decreases to be negative again with time. So the holographic EoP can exist only in some time interval. When $l<0.948$, the HMI is always zero. In the right panel, similar phenomenon is observed. We will show the evolution of holographic EoP in Fig. \ref{fig:VaidyaEoPtime}.

\begin{figure}
\begin{centering}
\includegraphics[scale=0.40]{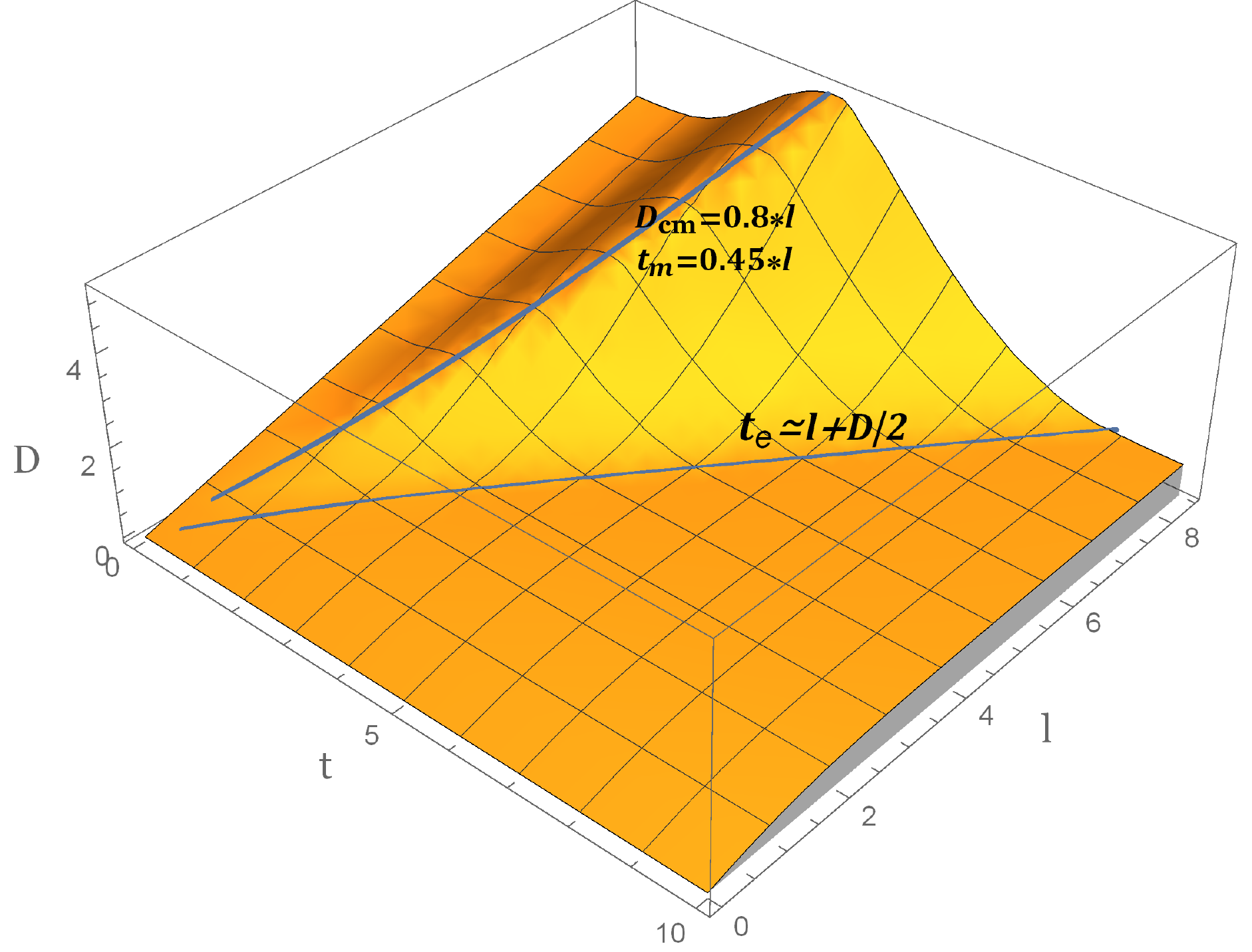}
\par\end{centering}
\caption{\label{fig:VaidyaEoPRegion}The region below the surface has non-vanishing holographic
EoP for two strips both with width $l$ separated by $D$ when
$d=2$. The maximum separation for given $l$ during the evolution is about $D_{cm}\simeq0.8l$ and the corresponding time $t_{m}\simeq0.45l$. The equilibrium time is about $t_e\simeq D/2+l$. For higher dimensional case, we  obtain similar qualitative behaviors. }
\end{figure}

The region allowing non-vanishing holographic EoP when $d=2$ is
shown in Fig. \ref{fig:VaidyaEoPRegion}. When $t<0$, the spacetime is pure $AdS_3$. The corresponding HMI (\ref{eq:VaidyaEoPCondition}) can be worked out
analytically,
\begin{equation}
I(l,D,t<0)=\frac{1}{2}\log\left(\frac{l^2}{D(D+2l)}\right).
\end{equation}
The critical separation for given $l$ is $D_c = (\sqrt{2}-1)l$.
When $t\to\infty$, the critical separation for given $l$ coincides with (\ref{DLBTZ1}) for three dimensional SAdS black brane which is shown in the left panel of Fig. \ref{fig:FiniteCriticalLentgth}. Note that no matter how large
$l$ is, the critical separation when $t\to\infty$ is $D_{c}(2,l)\leq D_c(2,\infty)=\ln2$.

When $l$ is fixed, the separation $D$ allowing holographic EoP increases
at first and then decreases as time $t$ grows. It can be shown that
the time $t_{m}$ needed to reach the maximum critical separation
$D_{cm}$ during the evolution has relation to $l$ by $t_{m}\simeq0.45l$
and $D_{cm}\simeq0.8l$ when $l$ is large enough. The time $t_{e}$
 needed to reach equilibrium is about $t_{e}\simeq l+D/2$. This can
be explained as follows. The HRT surface of strip $D$ reaches
equilibrium more early than that of strip $2l+D$. So $t_{e}$
can be approximated by the time which is needed so that $S_{2l+D}(t)$
reaches equilibrium. In asymptotic $AdS_{3}$ black brane, this time
is about $D/2+l$. In fact, it has been shown \cite{Cardy:2014rqa}
that for a given quench in 2D CFT, the density matrix of a strip with
width $L$ will be exponentially close to a thermal density matrix
if the time is larger than $L/2$. So the corresponding HRT surface
will reach equilibrium in time about $L/2$.

Similar behaviors are observed when $d>2$.  For pure $AdS_{d+1}$ spacetime,
the corresponding HMI when $t<0$ is
\begin{equation}
I(l,D,t<0)=\frac{2^{d-3}\pi^{\frac{d-1}{2}}V_{d-2}}{d-2}
\left(\frac{\Gamma(\frac{1}{2d-2})}{\Gamma(\frac{d}{2d-2})}\right)^{1-d}
\left(\frac{2}{l^{d-2}}-\frac{1}{D^{d-2}}-\frac{1}{(2l+D)^{d-2}}\right).
\end{equation}
The critical separation for given $l$  is still proportional to $l$. The coefficient can be worked out from above formula.
When $t\to\infty$, the critical separation for given $l$ is shown in  Fig. \ref{fig:FiniteCriticalLentgth}.
 In the evolution process, the time $t_{m}$ needed to reach the maximum critical separation
$D_{cm}$ and the equilibrium time $t_e$ still depend linearly on $l$. Only the coefficients
are relevant to dimension $d$. For example, when $d=3$, we have $t_m\simeq0.7l,D_{cm}\simeq0.9l,t_e\simeq0.66(2l+D)$.

Once (\ref{eq:VaidyaEoPCondition}) is satisfied, the holographic EoP is proportional
to the area of the extremal surface $\Gamma$ connecting $m$ and $m'$, as shown
in Fig. \ref{fig:HRT}. Due to the symmetry, the extremal surface lies
in the $(z,v)$ plane. The induced metric on this plane is
\begin{equation}
ds^{2}=\frac{1}{z^{2}}\left[-f(v,z)-2\frac{dz}{dv}\right]dv^{2}+\frac{1}{z^{2}}\sum_{i=1}^{d-2}dy_{i}^{2}.
\end{equation}
We can get the equation describing the extremal surface as
\begin{align}
0= & 2(d-1)f^{2}+4(d-1)z'^{2}-3zz'\partial_{z}f+f\left[6(d-1)z'-z\partial_{z}f\right]-z(2z''+\partial_{v}f).
\end{align}
 Here $z'=\frac{dz}{dv}$. Suppose the solution between $m=(z_{D},v_{D})$
and $m'=(z_{2l+D},v_{2l+D})$ can be expressed as $\tilde{z}(v)$.
The holographic EoP between the two strips $A$ and $B$ both with width $l$ separated by
$D$ is
\begin{align}
\frac{4 }{V_{d-2}}E(l,D,t)= & \int_{v_{D}(t)}^{v_{2l+D}(t)}\frac{1}{\tilde{z}^{d-1}}\sqrt{-f(v,\tilde{z})-2\frac{d\tilde{z}}{dv}}dv.
\end{align}
 It must be ensured that $v_{D}$ and $v_{2l+D}$ correspond
to the same boundary time $t$.

We show the evolution of holographic EoP for given $l$ and $D$ when $d=2$
in Fig. \ref{fig:VaidyaEoPtime}. The behaviors of holographic EoP in higher dimension are qualitatively similar. In the left panel, we fix the separation
$D=0.4$ and let $l$ run. In the right panel, we fix $l=1.5$ and let
$D$ run. When $l$ is large enough or $D$ is small enough, the holographic EoP for given $l$ and $D$
grows with time at first, and then decreases with time. The equilibrium time of holographic EoP is
\begin{equation}
\text{EoP:}~~t_e\approx l/2\,,
\end{equation}
which is almost independent of separation $D$. See the right panel of Fig.~\ref{fig:VaidyaEoPtime} for example. On the other hand, from Fig.~\ref{fig:VaidyaEoPRegion} and the right panel of Fig.~\ref{fig:VaidyaHMItime} we can conclude that the equilibrium time of HMI is
\begin{equation}
\text{HMI:}~~t_e\approx l+D/2\,,
\end{equation}
which depends on separation $D$. Thus, from the view of point of equilibrium time, we may conclude that MI is sensitive to the whole subsystem including strips $A,B$ and the separation, while EoP is only sensitive to strips $A$ and $B$ themselves. This behavior is more obvious in the right panel of Fig.~\ref{fig:VaidyaEoPtime}. We see that the equilibrium time is almost irrelevant to the separation.

In the left panel, when $l>1.12$, there is always nonvanishing holographic EoP in the whole evolution process. When $1.12>l>0.963$, the  holographic EoP is positive at first and then drops to zero at some critical time. When $0.963>l>0.948$, the  holographic EoP is zero at first, then jumps to be positive for some time and drops down to zero again at some critical time. When $l<0.948$, there is no  holographic EoP all the time. Similar behaviors are observed in the right panel.

\section{Summaries and discussions}\label{sum}
In this paper, we studied the holographic entanglement of purification for Schwarzschild-AdS  black branes and Vaidya-AdS black branes. For Schwarzschild-AdS  black branes, we considered two disjoint  strips with the same width on the same boundary and two boundaries respectively. For Vaidya-AdS black branes, we studied two disjoint strips with the same width on the same boundary.

For two disjoint trips on the same boundary of different dimensional Schwarzschild-AdS  black branes, we found that there are critical separations beyond which the  holographic EoP will vanish. When the strip width is small, the critical separation is linearly proportional to the strip width in which the coefficient depends on the spacetime dimension. When the strip width is very large, the critical separation is almost independent of the strip width, but inversely proportional to the spacetime dimension. For three dimensional black brane, the relationship between critical separation and strip width is given by Eq.~\eqref{DLBTZ1}. There are no compact analytical results in higher dimensions.
For fixed strip width, the  holographic EoP diverges when the separation goes to zero. As the separation grows, the EoP takes a nosedive. When the separation goes beyond the critical separation, the  holographic EoP drops discontinuously to zero. For fixed separation, the  holographic EoP vanishes when the strip width is small. It becomes positive discontinuously when the strips are wide enough. When the strip width is very large, the  holographic EoP tends to a saturation value. The larger the separation is, the smaller saturation of  holographic EoP.


In the case that the two strips lay symmetrically on the two-copy boundaries of the maximally extended Schwarzschild-AdS  black brane, we studied how the  holographic EoP evolves with respective to one boundary time $t_B$. When the strip width $l\rightarrow\infty$, the initial  holographic EoP is given by Eq.~\eqref{Egammat}, which is always nonzero. In the case that $d=2$, the growth behaviors of  holographic EoP and  holographic entanglement entropy are similar and show the linear growth at the late time limit. However, in the cases of $d>2$, the  holographic EoP will still grow with respective to time but the growth rate will exponentially decay to zero. If the width $l$ of strips is finite, there is a critical width $l_c$ and critical time $t_c(l)$ and the  holographic EoP is nonzero only when $l>l_c$ and $t_B<t_c(l)$. In this case, the  holographic EoP will first increase with respective to $t_B$ and suddenly drop into zero when $t_B \geq t_c(l)$.


We also considered the evolution of EoP after a thermal quench for CFT. This process can be described holographically by the Vaidya-AdS spacetime. We find that the critical separation allowing nonvanishing holographic EoP increases with time and then decreases to a smaller value at late time.  The maximum separation during the evolution is proportional to the strip width. The holographic EoP exists only when the HMI is positive. We find that when the strip width $l$ is large enough or the separation $D$ is small enough, the HMI is always positive. It grows with time at first but then decreases to a smaller value later. When $d=2$, the equilibrium time is about $l+D/2$. This can be understood from \cite{Cardy:2014rqa} which has shown that the reduced  density matrix of a strip with width $l$ towards to thermal equilibrium with time scale $l/2$  in two dimensional CFT. On the other hand, we find that the equilibrium time of  holographic EoP is about $l/2$ and is almost independent of the separation $D$. Thus we conclude that MI is sensitive to the whole subsystem including strips and the separation, while the EoP is only sensitive to strips themselves. When the strip width is small or the separation is large, the EoP changes discontinuously while the HMI changes continuously. Similar behaviors are found in higher dimensional spacetime.

In this paper, our discussion is limited to the leading order. The HMI of two disjoint region suffers a phase transition from nonzero to zero when the separation is larger than a critical distance \cite{Headrick:2010zt}.
However, the quantum mutual information satisfies  actually an inequality \cite{Wolf:2007tdq},
\begin{equation}
I(A, B)\ge \frac{\mathcal{C}(M_A, M_B)^2}{2||M_A||^2||M_B||^2}\,,
\end{equation}
where $M_A$ and $M_B$ are the observables in the regions $A$ and $B$ respectively, and $\mathcal{C}(M_A, M_B):=\langle M_A\otimes M_B \rangle-\langle M_A\rangle \langle M_B \rangle$ is the correlation function of $M_A$ and $M_B$. This indicates that, due to the quantum correlations, the quantum MI of two disjoint regions is usually not vanishing, even when they are far apart. Thus, when considering the quantum correlations, as a result of (\ref{ineqEoPS}), we should not expect the EoP disappears immediately after the transition point.
How to describe the quantum correction in the dual bulk is still a question.

\section*{Acknowledgements}
 W.-M. Li are supported in part by NSFC Grant No. 11275010, No. 11325522, No. 11335012 and No. 11735001. C.-Y. Zhang is supported by National Postdoctoral Program for Innovative Talents BX201600005 and Project funded by China Postdoctoral Science Foundation.

\bibliographystyle{JHEP}

\begin{thebibliography}{10}

\bibitem{Maldacena:1997re}
J.~M. Maldacena, \emph{{The Large N limit of superconformal field theories and
  supergravity}}, \href{http://dx.doi.org/10.1023/A:1026654312961,
  10.4310/ATMP.1998.v2.n2.a1}{\emph{Int. J. Theor. Phys.} {\bf 38} (1999)
  1113--1133}, [\href{http://arxiv.org/abs/hep-th/9711200}{{\tt
  hep-th/9711200}}].

\bibitem{PhysRevLett.96.181602}
S.~Ryu and T.~Takayanagi, \emph{Holographic derivation of entanglement entropy
  from the anti--de sitter space/conformal field theory correspondence},
  \href{http://dx.doi.org/10.1103/PhysRevLett.96.181602}{\emph{Phys. Rev.
  Lett.} {\bf 96} (May, 2006) 181602}.

\bibitem{Ryu:2006ef}
S.~Ryu and T.~Takayanagi, \emph{{Aspects of Holographic Entanglement Entropy}},
  \href{http://dx.doi.org/10.1088/1126-6708/2006/08/045}{\emph{JHEP} {\bf 08}
  (2006) 045}, [\href{http://arxiv.org/abs/hep-th/0605073}{{\tt
  hep-th/0605073}}].

\bibitem{PhysRevE.69.066138}
A.~Kraskov, H.~St\"ogbauer and P.~Grassberger, \emph{Estimating mutual
  information}, \href{http://dx.doi.org/10.1103/PhysRevE.69.066138}{\emph{Phys.
  Rev. E} {\bf 69} (Jun, 2004) 066138}.

\bibitem{Fischler:2012uv}
W.~Fischler, A.~Kundu and S.~Kundu, \emph{{Holographic Mutual Information at
  Finite Temperature}},
  \href{http://dx.doi.org/10.1103/PhysRevD.87.126012}{\emph{Phys. Rev.} {\bf
  D87} (2013) 126012}, [\href{http://arxiv.org/abs/1212.4764}{{\tt
  1212.4764}}].

\bibitem{Morrison:2012iz}
I.~A. Morrison and M.~M. Roberts, \emph{{Mutual information between
  thermo-field doubles and disconnected holographic boundaries}},
  \href{http://dx.doi.org/10.1007/JHEP07(2013)081}{\emph{JHEP} {\bf 07} (2013)
  081}, [\href{http://arxiv.org/abs/1211.2887}{{\tt 1211.2887}}].

\bibitem{Takayanagi:2017knl}
T.~Takayanagi and K.~Umemoto, \emph{{Holographic Entanglement of
  Purification}},
  \href{http://dx.doi.org/10.1038/s41567-018-0075-2}{\emph{Nature Phys.} {\bf
  14} (2018) 573--577}, [\href{http://arxiv.org/abs/1708.09393}{{\tt
  1708.09393}}].

\bibitem{doi:10.1063/1.1498001}
B.~M. Terhal, M.~Horodecki, D.~W. Leung and D.~P. DiVincenzo, \emph{The
  entanglement of purification},
  \href{http://dx.doi.org/10.1063/1.1498001}{\emph{Journal of Mathematical
  Physics} {\bf 43} (2002) 4286--4298}.

\bibitem{PhysRevA.91.042323}
S.~Bagchi and A.~K. Pati, \emph{Monogamy, polygamy, and other properties of
  entanglement of purification},
  \href{http://dx.doi.org/10.1103/PhysRevA.91.042323}{\emph{Phys. Rev. A} {\bf
  91} (Apr, 2015) 042323}.

\bibitem{Wolf:2007tdq}
M.~M. Wolf, F.~Verstraete, M.~B. Hastings and J.~I. Cirac, \emph{{Area Laws in
  Quantum Systems: Mutual Information and Correlations}},
  \href{http://dx.doi.org/10.1103/PhysRevLett.100.070502}{\emph{Phys. Rev.
  Lett.} {\bf 100} (2008) 070502}, [\href{http://arxiv.org/abs/0704.3906}{{\tt
  0704.3906}}].

\bibitem{Chen:2017hbk}
B.~Chen, L.~Chen, P.-x. Hao and J.~Long, \emph{{On the Mutual Information in
  Conformal Field Theory}},
  \href{http://dx.doi.org/10.1007/JHEP06(2017)096}{\emph{JHEP} {\bf 06} (2017)
  096}, [\href{http://arxiv.org/abs/1704.03692}{{\tt 1704.03692}}].

\bibitem{Chen:2017yns}
B.~Chen, Z.-Y. Fan, W.-M. Li and C.-Y. Zhang, \emph{{Holographic Mutual
  Information of Two Disjoint Spheres}},
  \href{http://dx.doi.org/10.1007/JHEP04(2018)113}{\emph{JHEP} {\bf 04} (2018)
  113}, [\href{http://arxiv.org/abs/1712.05131}{{\tt 1712.05131}}].

\bibitem{Bhattacharyya:2018sbw}
A.~Bhattacharyya, T.~Takayanagi and K.~Umemoto, \emph{{Entanglement of
  Purification in Free Scalar Field Theories}},
  \href{http://dx.doi.org/10.1007/JHEP04(2018)132}{\emph{JHEP} {\bf 04} (2018)
  132}, [\href{http://arxiv.org/abs/1802.09545}{{\tt 1802.09545}}].

\bibitem{Hirai:2018jwy}
H.~Hirai, K.~Tamaoka and T.~Yokoya, \emph{{Towards Entanglement of Purification
  for Conformal Field Theories}},
  \href{http://dx.doi.org/10.1093/ptep/pty063}{\emph{PTEP} {\bf 2018} (2018)
  063B03}, [\href{http://arxiv.org/abs/1803.10539}{{\tt 1803.10539}}].

\bibitem{2017arXiv171101288H}
J.~{Hauschild}, E.~{Leviatan}, J.~H. {Bardarson}, E.~{Altman}, M.~P. {Zaletel}
  and F.~{Pollmann}, \emph{{Finding purifications with minimal entanglement}},
  {\emph{ArXiv e-prints} (Nov., 2017) },
  [\href{http://arxiv.org/abs/1711.01288}{{\tt 1711.01288}}].

\bibitem{2018arXiv180100142C}
B.-B. {Chen}, L.~{Chen}, Z.~{Chen}, W.~{Li} and A.~{Weichselbaum},
  \emph{{Energy Scales and Exponential Speedup in Thermal Tensor Network
  Simulations}}, {\emph{ArXiv e-prints} (Dec., 2018) },
  [\href{http://arxiv.org/abs/1801.00142}{{\tt 1801.00142}}].

\bibitem{Nguyen:2017yqw}
P.~Nguyen, T.~Devakul, M.~G. Halbasch, M.~P. Zaletel and B.~Swingle,
  \emph{{Entanglement of purification: from spin chains to holography}},
  \href{http://dx.doi.org/10.1007/JHEP01(2018)098}{\emph{JHEP} {\bf 01} (2018)
  098}, [\href{http://arxiv.org/abs/1709.07424}{{\tt 1709.07424}}].

\bibitem{Bao:2017nhh}
N.~Bao and I.~F. Halpern, \emph{{Holographic Inequalities and Entanglement of
  Purification}}, \href{http://dx.doi.org/10.1007/JHEP03(2018)006}{\emph{JHEP}
  {\bf 03} (2018) 006}, [\href{http://arxiv.org/abs/1710.07643}{{\tt
  1710.07643}}].

\bibitem{Bao:2018gck}
N.~Bao and I.~F. Halpern, \emph{{Conditional and Multipartite Entanglements of
  Purification and Holography}},  \href{http://arxiv.org/abs/1805.00476}{{\tt
  1805.00476}}.

\bibitem{Espindola:2018ozt}
R.~Esp¨ªndola, A.~Guijosa and J.~F. Pedraza, \emph{{Entanglement Wedge
  Reconstruction and Entanglement of Purification}},
  \href{http://dx.doi.org/10.1140/epjc/s10052-018-6140-2}{\emph{Eur. Phys. J.}
  {\bf C78} (2018) 646}, [\href{http://arxiv.org/abs/1804.05855}{{\tt
  1804.05855}}].

\bibitem{Umemoto:2018jpc}
K.~Umemoto and Y.~Zhou, \emph{{Entanglement of Purification for Multipartite
  States and its Holographic Dual}},
  [\href{http://arxiv.org/abs/1805.02625}{{\tt 1805.02625}}].

\bibitem{Tamaoka:2018ned}
T.~Kotaro, \emph{{Entanglement Wedge Cross Section from the Dual Density Matrix}},
  [\href{http://arxiv.org/abs/1809.09109}{{\tt 1809.09109}}].

\bibitem{Headrick:2014cta}
M.~Headrick, V.~E. Hubeny, A.~Lawrence and M.~Rangamani, \emph{{Causality \&
  holographic entanglement entropy}},
  \href{http://dx.doi.org/10.1007/JHEP12(2014)162}{\emph{JHEP} {\bf 12} (2014)
  162}, [\href{http://arxiv.org/abs/1408.6300}{{\tt 1408.6300}}].

\bibitem{Maldacena:2001kr}
J.~M. Maldacena, \emph{{Eternal black holes in anti-de Sitter}},
  \href{http://dx.doi.org/10.1088/1126-6708/2003/04/021}{\emph{JHEP} {\bf 04}
  (2003) 021}, [\href{http://arxiv.org/abs/hep-th/0106112}{{\tt
  hep-th/0106112}}].

\bibitem{Hartman:2013qma}
T.~Hartman and J.~Maldacena, \emph{{Time Evolution of Entanglement Entropy from
  Black Hole Interiors}},
  \href{http://dx.doi.org/10.1007/JHEP05(2013)014}{\emph{JHEP} {\bf 05} (2013)
  014}, [\href{http://arxiv.org/abs/1303.1080}{{\tt 1303.1080}}].

\bibitem{MIyaji:2015mia}
M.~Miyaji, T.~Numasawa, N.~Shiba, T.~Takayanagi and K.~Watanabe,
  \emph{{Distance between Quantum States and Gauge-Gravity Duality}},
  \href{http://dx.doi.org/10.1103/PhysRevLett.115.261602}{\emph{Phys. Rev.
  Lett.} {\bf 115} (2015) 261602}, [\href{http://arxiv.org/abs/1507.07555}{{\tt
  1507.07555}}].

\bibitem{Sinamuli:2016rms}
M.~Sinamuli and R.~B. Mann, \emph{{Geons and the Quantum Information Metric}},
  \href{http://dx.doi.org/10.1103/PhysRevD.96.026014}{\emph{Phys. Rev.} {\bf
  D96} (2017) 026014}, [\href{http://arxiv.org/abs/1612.06880}{{\tt
  1612.06880}}].

\bibitem{Chen:2018mcc}
B.~Chen, W.-M. Li, R.-Q. Yang, C.-Y. Zhang and S.-J. Zhang, \emph{{Holographic
  subregion complexity under a thermal quench}},
  \href{http://dx.doi.org/10.1007/JHEP07(2018)034}{\emph{JHEP} {\bf 07} (2018)
  034}, [\href{http://arxiv.org/abs/1803.06680}{{\tt 1803.06680}}].

\bibitem{Allais:2011ys}
A.~Allais and E.~Tonni, \emph{{Holographic evolution of the mutual
  information}}, \href{http://dx.doi.org/10.1007/JHEP01(2012)102}{\emph{JHEP}
  {\bf 01} (2012) 102}, [\href{http://arxiv.org/abs/1110.1607}{{\tt
  1110.1607}}].

\bibitem{Ziogas:2015aja}
V.~Ziogas, \emph{{Holographic mutual information in global Vaidya-BTZ
  spacetime}}, \href{http://dx.doi.org/10.1007/JHEP09(2015)114}{\emph{JHEP}
  {\bf 09} (2015) 114}, [\href{http://arxiv.org/abs/1507.00306}{{\tt
  1507.00306}}].

\bibitem{Cardy:2014rqa}
J.~Cardy, \emph{{Thermalization and Revivals after a Quantum Quench in
  Conformal Field Theory}},
  \href{http://dx.doi.org/10.1103/PhysRevLett.112.220401}{\emph{Phys. Rev.
  Lett.} {\bf 112} (2014) 220401}, [\href{http://arxiv.org/abs/1403.3040}{{\tt
  1403.3040}}].

\bibitem{Headrick:2010zt}
M.~Headrick, \emph{{Entanglement Renyi entropies in holographic theories}},
  \href{http://dx.doi.org/10.1103/PhysRevD.82.126010}{\emph{Phys. Rev.} {\bf
  D82} (2010) 126010}, [\href{http://arxiv.org/abs/1006.0047}{{\tt
  1006.0047}}].

\end{thebibliography}

\end{document}